\documentclass[prd,aps,preprint,floatfix,nofootinbib,superscriptaddress]{revtex4}
\usepackage{epsfig}

\usepackage[reqno]{amsmath}
\usepackage{array}
\usepackage[hypertex]{hyperref}
 
\def\gappeq{\mathrel{\rlap {\raise.5ex\hbox{$>$}}
{\lower.5ex\hbox{$\sim$}}}}
\def\lappeq{\mathrel{\rlap{\raise.5ex\hbox{$<$}}
{\lower.5ex\hbox{$\sim$}}}}
\def\CPV{\begin{picture}(20,0)(0,0)
        \put(0,0){CP}
        \put(0,0){\line(2,1){18}}
        \end{picture} }
\def\beq{\begin{equation}} \def\eeq{\end{equation}}
\def\bea{\begin{eqnarray}} \def\eea{\end{eqnarray}}
\def\bq{\begin{quote}} \def\eq{\end{quote}}

\input epsf.tex
\def\DESepsf(#1 width #2){\epsfxsize=#2 \epsfbox{#1}}
\usepackage{graphicx}

\begin{document}




\title{\Large Theory of Neutrinos}

\author{R.N.~Mohapatra (Group Leader)}
\affiliation{University of Maryland, College Park, MD 20742, USA}

\author{S.~Antusch}
\affiliation{University of Southampton, Southampton SO17 1BJ, U.K}

\author{K.S.~Babu}
\affiliation{Oklahoma State University, Stillwater, OK-74078, USA}

\author{G.~Barenboim}
\affiliation{University of Valencia, Valencia, Spain}

\author{Mu-Chun Chen}
\affiliation{Brookhaven National Laboratory, Upton, NY}

\author{S.~Davidson}
\affiliation{IPPP, University of Durham, Durham, DH1 3LE, Great Britain}

\author{A.~de Gouv\^ea}
\affiliation{Northwestern University, Evanston, IL-60208}

\author{P.~de Holanda}
\affiliation{Instituto de F\'\i sica Gleb Wataghin, UNICAMP
PO BOX 6165, CEP 13083-970, Campinas - SP, Brazil}

\author{B.~Dutta}
\affiliation{University of Regina, Canada}

\author{Y.~Grossman}
\affiliation{Technion--Israel Institute of Technology,
Technion City, 32000 Haifa, Israel}

\author{A.~Joshipura}
\affiliation{Physical Research Laboratory, Ahmedabad 380009, India}


\author{J.~Kersten}
\affiliation{Physik-Department, Technische Universit{\"a}t M{\"u}nchen,
85748 Garching, Germany}

\author{Y.~Y.~Keum}
\affiliation{Nagoya University, Japan}

\author{S.~F.~King}
\affiliation{University of Southampton, Southampton SO17 1BJ, U.K}

\author{P.~Langacker}
\affiliation{University of Pennsylvania, Philadelphia, Pa-19104}

\author{M.~Lindner}
\affiliation{Physik-Department, Technische Universit{\"a}t M{\"u}nchen,
85748 Garching, Germany}

\author{W.~Loinaz}
\affiliation{Amherst College, Amherst, Ma-01002}

\author{I.~Masina} 
\affiliation{Fermi Center, Via Panisperna 89/A, I-00184 Roma, Italy
  and INFN, Sezione di Roma, "La Sapienza" Univ., P.le A. Moro 2,
  I-00185 Roma, Italy}

\author{I.~Mocioiu}
\affiliation{University of Arizona, Tucson, AZ 85718, USA}

\author{S.~Mohanty}
\affiliation{Physical Research Laboratory, Ahmedabad 380009, India}

\author{H.~Murayama}\thanks{ On leave of absence from Department of
    Physics, University of California, Berkeley, CA 94720.}
\affiliation{School of Natural Sciences, Institute for Advanced Study,
  Princeton, NJ 08540, USA}

\author{S.~Pascoli}
\affiliation{UCLA, Los Angeles, CA 90095-1547, USA}

\author{S.~Petcov}
\affiliation{SISSA, Trieste, Italy}

\author{A.~Pilaftsis}
\affiliation{University of Manchester, Manchester M13 9PL, England}

\author{P.~Ramond}
\affiliation{University of Florida, Galinsville, Fa-32611}

\author{M.~Ratz}
\affiliation{Deutsches Elektronen-Synchrotron DESY, 22603 Hamburg,
Germany}

\author{W.~Rodejohann}
\affiliation{Scuola Internazionale Superiore di Studi Avanzati
Via Beirut 2-4, I-34014 Trieste}

\author{R. Shrock}
\affiliation{State University of New York at Stony Brook, NY}

\author{T.~Takeuchi}
\affiliation{Virginia Tech, Blacksburg, VA 24061}

\author{T.~Underwood}
\affiliation{University of Manchester, Manchester M13 9PL, England}

\author{F.~Vissani}
\affiliation{INFN, Laboratori Nazionali del Gran Sasso, Theory group,
  67010 Assergi (AQ), Italy}

\author{L.~Wolfenstein}
\affiliation{Carnegie-Mellon University, Pittsburgh, Pa-15213.}

\date{June, 2004}

\begin{abstract}
After a brief overview of the present knowledge of neutrino masses and
mixing, we summarize what can be learned about physics beyond the
standard model from the various proposed neutrino experiments. We also
comment on the impact of the experiments on our understanding of the
origin of the matter--antimatter asymmetry of the Universe as well as what
can be learned from some experiments outside the domain of neutrinos.
\end{abstract}

\maketitle

\tableofcontents
 
\section{Introduction}

Our understanding of neutrinos has changed tremendously in the past
six years.  Thanks to the efforts of several neutrino oscillation
studies of solar, atmospheric and reactor (anti)neutrinos, we learned
that neutrinos produced in a well defined flavor eigenstate can be
detected, after propagating a macroscopic distance, as a different
flavor eigenstate. The simplest interpretation of this phenomenon is
that, like all charged fermions the neutrinos have mass and that,
similar to quarks, the neutrino weak, or flavor, eigenstates are
different from neutrino mass eigenstates, i.e., neutrinos
mix \cite{BPont57}.  This new state of affairs has also raised many
other issues which did not exist for massless neutrinos: For example,
(i) massive neutrinos can have nonzero magnetic moments, like the
electron and the quarks; (ii) the heavier neutrinos may decay into
lighter ones, like charged leptons and quarks, and (iii) (most
importantly) the neutrinos can be either Majorana or Dirac fermions
\cite{barger}.

Learning about all these possibilities can not only bring our
knowledge of neutrinos to the same level as that of charged leptons
and quarks, but may also lead to a plethora of laboratory as well as
astrophysical and cosmological consequences with far reaching
implications. Most importantly, knowing neutrino properties in detail
may also play a crucial role in clarifying the blueprint of new
physical laws beyond those embodied in the Standard Model.

One may also consider the possibility that there could be new neutrino
species beyond the three known ones $(\nu_e,\nu_\mu, \nu_\tau)$. In
addition to being a question whose answer would be a revolutionary
milestone pointing to unexpected new physics, it may also become a
necessity if the LSND results are confirmed by the MiniBooNE
experiment, now in progress at Fermilab. This would, undoubtedly, be a
second revolution in our thinking about neutrinos and the nature of
unification.

The existence of neutrino masses qualifies as the first evidence of
new physics beyond the Standard Model. The answers to the
neutrino-questions mentioned above will add substantially to our
knowledge about the precise nature of this new physics, and in turn
about the nature of new forces beyond the Standard Model. They also
have the potential to unravel some of the deepest and long-standing
mysteries of cosmology and astrophysics, such as the origin of matter,
the origin of the heavy elements, and, perhaps, even the nature of
dark energy.

Active endeavors are under way to launch the era of precision neutrino
measurement science (PNMS), that will surely broaden the horizon of
our knowledge about neutrinos. We undertake this survey to pin down
how different experimental results expected in the coming decades can
elucidate the nature of neutrinos and our quest for new physics. In
particular, we would like to know (i) the implications of neutrinos
for such long standing ideas as grand unification, supersymmetry,
extra dimensions, etc; (ii) the implications of the possible existence
of additional neutrino species for physics and cosmology, and (iii)
whether neutrinos have anything to do with the origin of the observed
matter-antimatter asymmetry in the universe and, if so, whether there
is any way to determine this via low-energy experiments. Once the
answers to these questions are at hand, we will have considerably
narrowed the choices of new physics, providing a giant leap in our
understanding of the physical Universe.

The purpose of this document is to briefly summarize what we know
about neutrino masses and mixings, their context in overall physics,
their connection to theoretical models, and open questions.  There is
a companion document (``Theory White Paper'') where further technical
details are presented.  

\section{Why Neutrinos?}

Neutrinos are elementary particles with spin 1/2,
electrically neutral, and obey Fermi-Dirac statistics.  Even though
their existence has been known since the 1950s and the existence 
of three types of them was experimentally 
confirmed in the 1990s, it has been difficult to study their
intrinsic properties due to their weak interactions.  Nonetheless,
they have rather unique roles in the world of elementary particles.
They are ubiquitous in our universe, provide a unique window to
physics at very short distances, and may even be relevant to the
question ``Why do we exist?''  Moreover, the history of neutrinos has
been full of surprises, which is likely to continue in the future.

\subsection{Ubiquitous Neutrinos}

Neutrinos are the most ubiquitous matter particles in the universe.
They were produced in the Big Bang, when universe was so dense that
neutrinos, despite their only weak interactions, were in thermal
equilibrium with all other particle species.  Similarly to the cosmic
microwave background photons, their number density has been diluted by
the expansion of the universe.  In comparison, constituents of
ordinary matter, electrons, protons, and neutrons, are far rarer than
photons and neutrinos, by about a factor of ten billions.  It is clear
that we need to understand neutrinos in order to understand our
universe.


In terms of energy density, the yet unknown dark matter and dark energy
dominate the universe.  If neutrinos were massless, their energy
density could have been completely negligible in our current
universe.  However, in the last several years, we learned that neutrinos
have small but finite masses, implying that the neutrinos contribute to the total
energy of the universe at least as much as all stars combined.  We do not yet know the
mass of neutrinos precisely, and they may in fact be a
sizable fraction of dark matter.  The precise amount of the neutrino
component is relevant to the way galaxies and stars were formed during
the evolution of the universe.

Neutrinos are an important part of the stellar dynamics; without them,
stars would not shine.  There are about $7 \times 10^9 \:{\rm cm}^{-2}
\sec^{-1}$ neutrinos from the Sun reaching (and streaming through) the
Earth.  They also govern the dynamics of supernovae.

\subsection{Special Role of Neutrino Mass}

One way to characterize physics is an attempt to understand nature at
its most fundamental level, namely at the shortest distance scales, or
equivalently the highest energy scales, possible.  There has been two
approaches.  One way to access physics at the highest energy scales
possible is to build powerful particle accelerators and reach the
energy scale directly.  Another way is to look for rare effects from
physics at high-energy scales that do not occur from physics at known
energy scales, namely the Standard Model.  Physics of neutrino mass
(currently) belongs to the second category.

Rare effects from physics beyond the Standard Model are parameterized
by effective operators added to the Standard Model Lagrangian,
\begin{equation}
  {\cal L} = {\cal L}_{SM} + \frac{1}{\Lambda} {\cal L}_5
  + \frac{1}{\Lambda^2} {\cal L}_6 .
\end{equation}
The effects in ${\cal L}_5$ are suppressed by a single power of the
high energy scale, ${\cal L}_6$ by two powers, etc.  The possible terms
have been classified systematically by Weinberg, and there are
many terms suppressed by two powers:
\begin{equation}
  {\cal L}_6 \supset QQQL,\, \bar{L} \sigma^{\mu\nu} W_{\mu\nu} H e,\,
  W_\nu^\mu W_\lambda^\nu B_\mu^\lambda,\, 
  \bar{s}d \bar{s}d,\, (H^\dagger D_\mu H) (H^\dagger D^\mu H), \cdots\, .
\end{equation}
The examples here contribute to proton decay, $g-2$, the anomalous triple
gauge boson vertex, $K^0$--$\overline{K}^0$ mixing, and the
$\rho$-parameter, respectively.  It is interesting that there is only
one operator suppressed by a single power, ${\cal L}_5 = (LH) (LH)$.
After substituting the expectation value of the Higgs, the Lagrangian
becomes
\begin{equation}
  {\cal L} = \frac{1}{\Lambda} (LH)(LH)
  \rightarrow \frac{1}{\Lambda} (L\langle H\rangle)(L\langle H\rangle)
  = m_\nu \nu \nu,
  \label{eq:D5}
\end{equation}
nothing but the neutrino mass.

Therefore the neutrino mass plays a very unique role.  It is the
lowest-order effect of physics at short distances.  This is an
extremely small effect.  Any kinematical effects of the neutrino mass
are suppressed by $(m_\nu / E_\nu)^2$, and for $m_\nu \sim 1$~eV
(which we now know is already too large) and $E_\nu \sim 1$~GeV for
typical accelerator-based neutrino experiments, it is as small as
$(m_\nu / E_\nu)^2 \sim 10^{-18}$.  At first sight, there is no hope
to probe such a small number.  However, any physicist knows that
interferometry is a sensitive method to probe extremely small effects.
For interferometry to work, we need a coherent source.  Fortunately
there are many coherent sources of neutrinos: the Sun, cosmic rays,
reactors, etc.  We also need interference for an interferometer to
work.  Fortunately, there are large mixing angles that make the
interference possible.  We also need long baselines to enhance the
tiny effects. Again fortunately there are many long baselines
available, such as the size of the Sun, the size of the Earth, etc.
nature was very kind to provide all the necessary conditions for
interferometry to us!  Neutrino interferometry, a.k.a.  neutrino
oscillation, is a unique tool to study physics at very high energy
scales.

At the currently accessible energy scale of about a hundred GeV in
accelerators, the electromagnetic, weak, and strong forces have very
different strengths.  But their strengths become the same at $2\times
10^{16}$~GeV if there the Standard Model is extended to become
supersymmetric.  Given this, a natural candidate energy scale for new
physics is $\Lambda \sim 10^{16}$~GeV, which suggests $m_\nu \sim
\langle H \rangle^2/\Lambda \sim 0.003$~eV.  Curiously,
the data suggest numbers
quite close to this expectation.  Therefore neutrino oscillation experiments
may be probing physics at the energy scale of grand unification.

\begin{figure}
  \centering
  \includegraphics[width=0.5\textwidth]{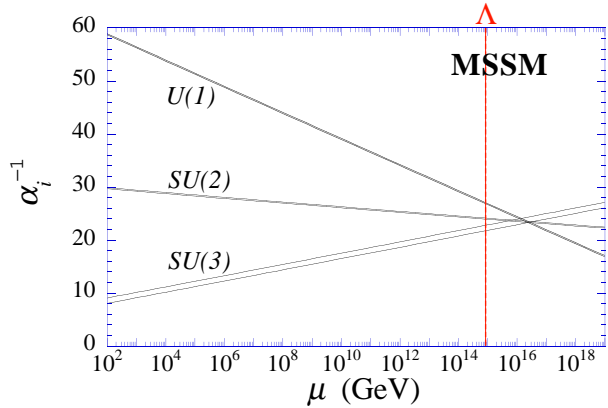}
  \vspace{-2mm}
  \caption{Apparent unification of gauge coupling unification in the
    MSSM at $2 \times 10^{16}$~GeV, compared to the suggested scale of
    new physics from the neutrino oscillation data.}\label{fig:MSSM}
\end{figure}

\subsection{Surprises}

Even though some may argue that the neutrino mass was observed with
the theoretically expected order of magnitude, it is fair to say that
we had not anticipated another important leptonic property: neutrinos
``oscillate'' from one species to another with a high probability.
Their mixing angles are large.  We've known that different species of
quarks mix, but their mixing angles are very small; the largest one is
13 degrees, and others are much smaller.  In comparison, the known
mixing of muon and tau neutrinos is consistent with being {\it
maximal}\/, or 45 degrees.  

Another surprise is the neutrino mass spectrum.  The quarks and charged
leptons have what are referred to as hierarchical mass spectra, namely that the
masses of charged fermions that share the same quantum numbers 
are drastically different among three generations of elementary particles.  
For instance, the masses of the
up- and top-quarks are different by almost five orders of magnitude.  On the
other hand, the two heavier neutrino masses differ {\it at most}\/ by a
factor of about five, and may possibly even be degenerate.  

Our knowledge of neutrino masses and mixing angles is still imprecise and 
incomplete.  Future measurements are likely to bring more surprises.


\subsection{Why Do We Exist?}

There is very little matter in our
universe.  That was not the case at an earlier stage.  There were
approximately equal amount of matter and anti-matter particles as
there were photons and neutrinos.  As the universe cooled, most of the matter
and anti-matter annihilated into pure energy, and disappeared.
The fact that there is still matter left means that there must have
been a small imbalance between matter and anti-matter, at the level of
one part in ten billion.  If not for this excess, all matter would 
have annihilated with all the anti-matter and we would not exist.  Why do
we exist?  The scientific question is rather what caused this tiny imbalance.  It
turns out that the finite mass of the neutrino may well have played a fundamental role.

All neutrinos detected are left-handed, namely that their spins point
the opposite directions from their momenta.  Likewise, all
anti-neutrinos are right-handed.  

Now that neutrinos were found to have finite masses, as discussed in
the previous section, we have to incorporate the massive neutrinos by
extending the Standard Model.  Note that massive neutrinos do not
travel at speed of light.  In principle, an observer can go faster
than the neutrino and look back at it.  Then a neutrino would appear
right-handed to the observer.  This is a state we have not seen
before.  Is this a new particle?  If so, we have to introduce
right-handed neutrinos, which do not have any of the Standard Model
gauge interactions, into the theory.  This is the possibility of the
Dirac neutrino.  On the other hand, we already know neutral
right-handed fermions: anti-neutrinos.  Could this state be an
anti-neutrino?  If so, we have to abandon the fundamental distinction
between neutrinos and anti-neutrinos, and hence matter and
anti-matter.  This is the possibility of the Majorana neutrino.

If the neutrino is a Majorana fermion, the the neutrino and the
anti-neutrino are the same
object.  This being the case,  it becomes possible for matter to transform into
anti-matter and vice-versa.  Therefore, the existence of neutrino masses
makes it possible
to create an imbalance between matter and anti-matter in early
universe.  This possibility is known as ``leptogenesis.''  In other
words, neutrino masses may play a role in providing the environment
necessary for our very existence.%
\footnote{It has been shown, however, that leptogenesis is
  possible also for Dirac neutrinos (see Subsec.~\ref{sec:DiracLeptogenesis}).}

\section{Our present knowledge about masses and mixings}

\subsection{Dirac versus Majorana Neutrinos}

The fact that the neutrino has no electric charge endows it with
certain properties not shared by the charged fermions of the Standard
Model: i.e.\ it can be its own antiparticle without violating electric
charge conservation. In that case, the neutrino is called a Majorana
fermion; otherwise it is called a Dirac neutrino. This leads to a
whole new class of experimental signatures, the most prominent among
them being the process of neutrinoless double beta decay of heavy
nuclei, ($\beta\beta_{0\nu}$). Since $\beta\beta_{0\nu}$ arises due to
the presence of neutrino Majorana masses, the observation of
$\beta\beta_{0\nu}$ decay, in addition to establishing the existence
of lepton number violation, can also provide very precise information
about neutrino masses and mixing, provided (i) one can satisfactorily
eliminate other contributions to this process that may arises from
other interactions in full beyond-the-Standard-Model theory, as we
discuss below, (ii) one can precisely estimate the values of the
nuclear matrix elements associated with the $\beta\beta_{0\nu}$ in
question.

\subsection{Neutrino masses and mixings}
We will use the notation where the
weak-eigenstates (defined as the neutrino that is produced in a
charged-current weak
interaction process associated with a well-defined charged lepton) are
denoted by
$\nu_{\alpha}$ (with $\alpha~=~e, {\mu}, {\tau},\ldots$), where the
ellipsis indicate yet to be 
discovered ``sterile" states
that are not produced in association with charged leptons and/or
fourth-generation neutrinos. 

Let us now focus on the case of only three Majorana neutrinos, with
mass matrix $m_{\nu}^{\alpha\beta}$ in the weak-eigenbasis (note that
$m_{\nu}$ is symmetric, i.e.,
$m_{\nu}^{\alpha\beta}=m_{\nu}^{\beta\alpha}$).  In the weak-basis
where the charged lepton mass-matrix and the charged current
coupling-matrix is diagonal, the neutrino mass-matrix is
\begin{equation}
  m^{\alpha\beta}_{\nu}=\sum_i(U^{*})_{\alpha i}m_i (U^{\dagger})_{i\beta},
  \label{m_ab}
\end{equation}
where $U$ is the Maki-Nakagawa-Sakata-Pontecorvo (MNSP) matrix, and $m_i$,
$i=1,2,3$, are the neutrino mass-eigenvalues, which can be taken real and
positive. We choose to write $U=V\times K$, where $V$ and $K$ are given by
Eq.~(\ref{V}).  The phases in $K$ are the so-called Majorana phases.  The case
for Dirac neutrinos is similar, except that $K$ can be absorbed into the phases
of neutrino mass eigenstates.

The mass-eigenstate $\nu_i$, $i=1, 2, 3,\ldots$, has a well-defined mass $m_i$
and we will order the mass eigenvalues such that $m_1^2<m^2_2$ and $\Delta
m^2_{12}<|\Delta m^2_{13}|$, where $\Delta m^2_{ij}\equiv m_j^2-m_i^2$. Flavor
eigenstates are expressed in terms of the mass eigenstates as follows:
$\nu_{\alpha}=\sum_i U_{\alpha i}\nu_i$. $U_{\alpha i}$ are the elements of the
MNSP matrix, and are related to the observable mixing angles in the basis where
the charged lepton masses are diagonal.

For the case of three Majorana neutrinos, the MNSP matrix $U$ can be
written as: $V K$, where $V$ will be parameterized as 
\begin{equation}
  \begin{pmatrix}
    c_{12}c_{13} & s_{12}c_{13} & s_{13} e^{-i\delta} \cr
    -s_{12}c_{23}-c_{12}s_{23}s_{13} e^{i\delta} &
    c_{12}c_{23}-s_{12}s_{23}s_{13} e^{i\delta}
    & s_{23}c_{13} \cr
    s_{12}s_{23}-c_{12}c_{23}s_{13} e^{i\delta} &
    -c_{12}s_{23}-s_{12}c_{23}s_{13} e^{i\delta} & c_{23}c_{13} 
  \end{pmatrix}, \quad
  K=\left(
    \begin{array}{ccc}
      1 & &\\ & e^{i\phi_1} &\\ & & e^{i\phi_2}
    \end{array} \right)\;.
  \label{V}
\end{equation}

\begin{figure}[tbp]
  \centering
  \includegraphics[height=12cm]{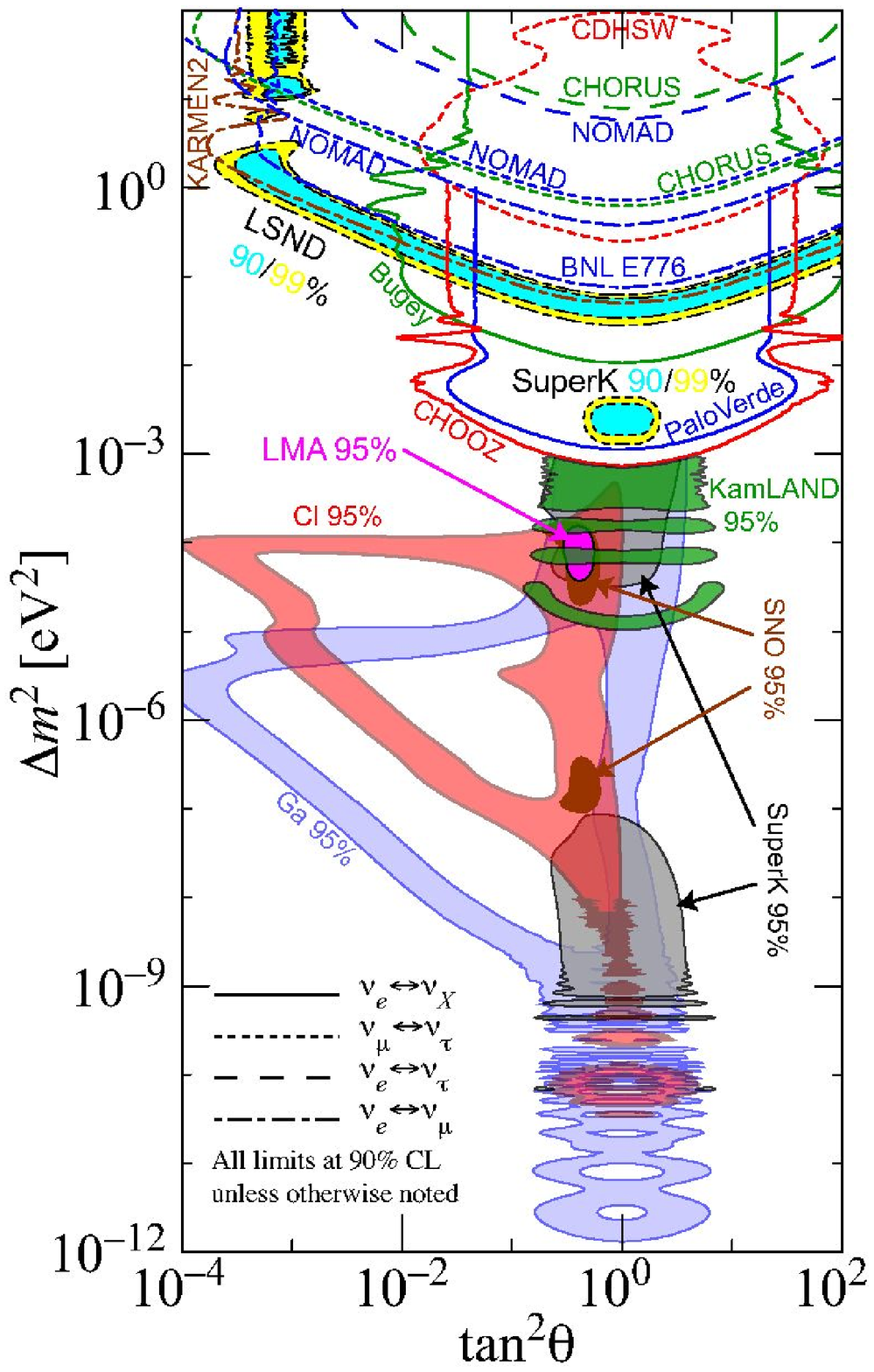}
  \includegraphics[height=12cm]{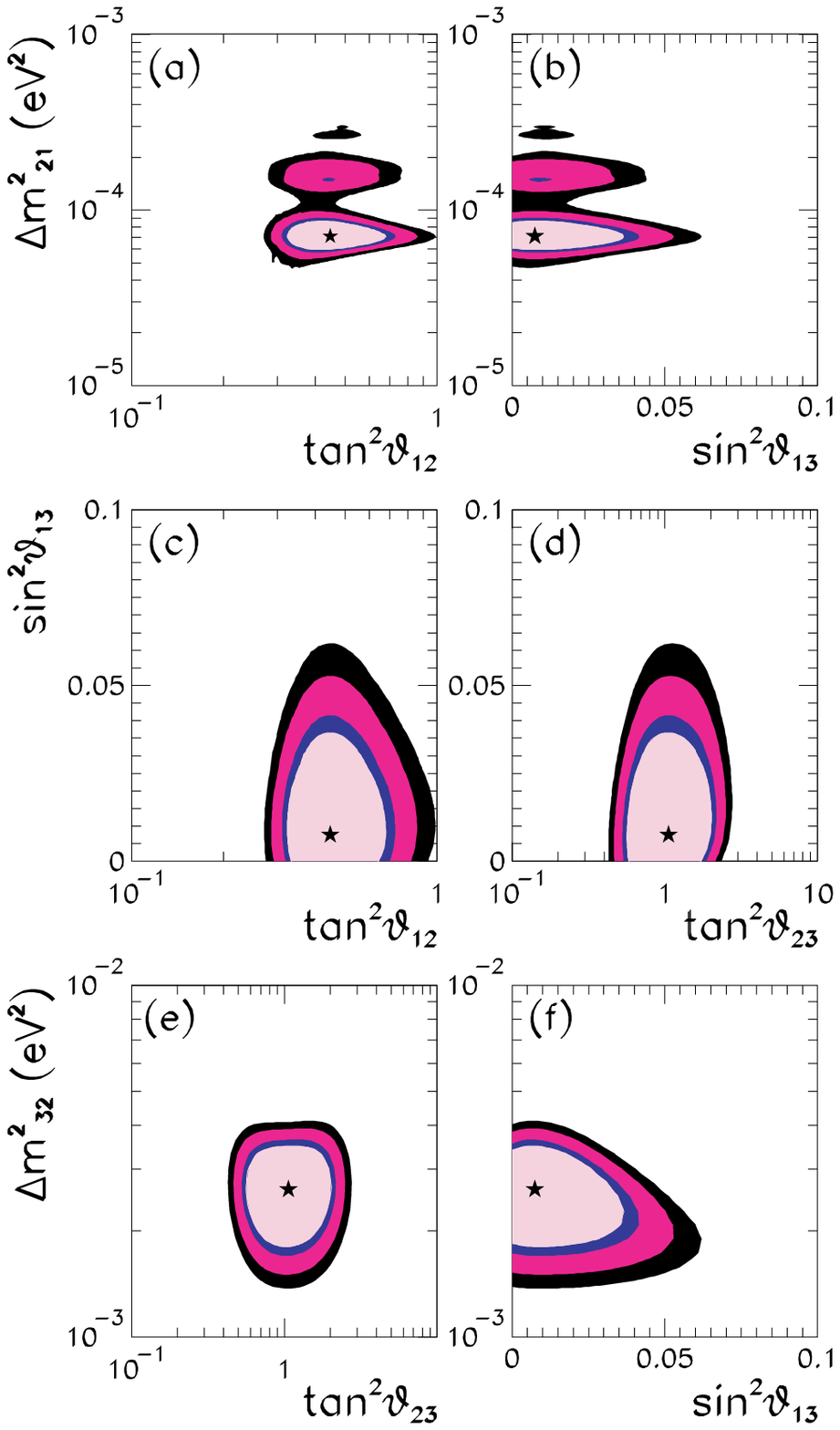}
  \caption{Left: Compilation of various neutrino oscillation
    experiments.  Right: Global fit to solar, atmospheric, and reactor
    neutrino oscillation data.}
  \label{fig:globalfit}
\end{figure}

Neutrino oscillation experiments have already provided measurements for  the
neutrino mass-squared differences, as well as the mixing angles. The 
allowed values for the $\theta_{ij}$ as well as the $\Delta m^2$'s are, at the
3$\sigma$ level: $\sin^22\theta_{23}\geq 0.92$; 
$1.2\times 10^{-3}~{\rm eV}^2
\leq|\Delta m^2_{13}| \leq 4.8\times 10^{-3}~{\rm eV}^2$; $0.70 \leq
\sin^22\theta_{12} \leq 0.95$; $5.4\times 10^{-5}~{\rm eV}^2 \leq \Delta m^2_{12}
\leq 9.5\times 10^{-5}~{\rm eV}^2$; $\sin\theta_{13}\leq 0.23$.
There is currently no constraint on any of the CP-odd phases or on the
sign of $\Delta m^2_{13}$.

\begin{figure}[tbp]
  \centering
  \includegraphics[width=0.5\textwidth]{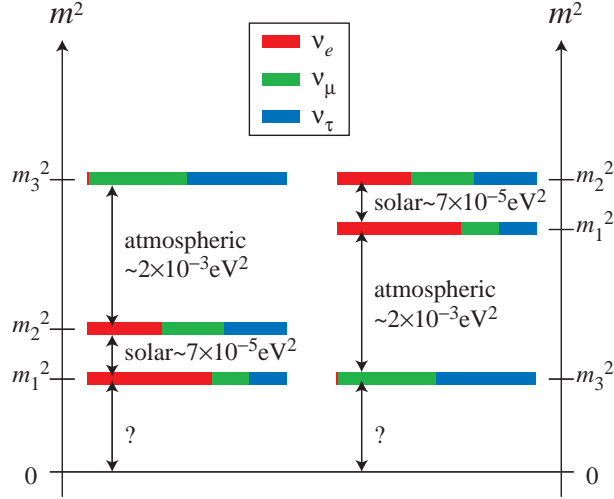}
  \caption{Neutrino masses and mixings as indicated by the current data.}
  \label{fig:neutrinospectrum}
\end{figure}

Since the oscillation data are only sensitive to mass-squared differences, they
allow for three possible arrangements of the different mass levels:
\renewcommand{\labelenumi}{(\roman{enumi})}
\begin{enumerate}
\item Normal hierarchy, i.e.\ $m_1\ll m_2 \ll m_3$. In this case,
we can deduce the value of $m_3 \simeq \sqrt{\Delta m^2_{23}}\simeq
0.03-0.07$ eV. In this case $\Delta m^2_{23}\equiv m^2_3-m^2_2 > 0$.
 The solar neutrino oscillation involves the two lighter levels. The mass
of the lightest neutrino is unconstrained. If $m_1\ll m_2$, then we get
the value of $m_2 \simeq 0.008$~eV.

\item Inverted hierarchy, i.e.\ $m_1 \simeq m_2 \gg m_3$ with
$m_{1,2} \simeq \sqrt{\Delta m^2_{23}}\simeq 0.03-0.07$ eV. In this case,
solar neutrino oscillation takes place between the heavier levels and we
have $\Delta m^2_{23}\equiv m^2_3-m^2_2 < 0$. We have no information about
$m_3$ except that its value is much less than the other two masses.

\item Degenerate neutrinos, i.e.\ $m_1\simeq m_2 \simeq m_3$.

\end{enumerate}

Oscillation experiments do not tell us about the overall scale of masses.  It
is therefore important to explore to what extent the absolute values of the
masses can be determined. While discussing the question of absolute masses, it
is good to keep in mind that none of the methods discussed below can provide
any information about the lightest neutrino mass in the cases of a normal or
inverted mass-hierarchy. They are most useful for determining absolute masses
in the case of degenerate neutrinos, i.e., when all $m_i\geq 0.1$~eV.

One can directly search for the kinematical effect of nonzero neutrino
masses in beta-decay by looking for structure near the end point
of the electron energy spectrum.  This search is sensitive to neutrino
masses regardless of whether the neutrinos are Dirac or Majorana
particles.  One is sensitive to the quantity $m_{\beta}\equiv
\sqrt{\sum_i |U_{ei}|^2m^2_i}$. The Troitsk and Mainz experiments
place the present upper limit on $m_\beta \leq 2.2$~eV.  The proposed
KATRIN experiment is projected to be sensitive to $m_{\beta}>0.2$~eV,
which will have important implications for the theory of neutrino
masses. For instance, if the result is positive, it will imply a
degenerate spectrum; on the other hand a negative result will be a
very useful constraint.

If neutrinos are Majorana particles, the rate for $\beta\beta_{0\nu}$
decay Majorana mass for the neutrino \cite{Valle} depends on the
combination $m_{ee} = \sum U^2_{ei} m_i$, provided heavy particle
contributions to this process present in various theories are
small \cite{moh1}. If they are not small however (which will need
additional experiments to decide), observing $\beta\beta_{0\nu}$ decay
will still be of fundamental significance since it will provide the
first observation of lepton number violation, for which there is
strong theoretical motivation.

The present best upper bound on $\beta\beta_{0\nu}$ decay lifetimes
come from the Heidelberg-Moscow and the IGEX experiments and can be
translated into an upper limit on $m_{ee}\lesssim 0.3 $~eV. There is a
claim of discovery of neutrinoless double beta decay of enriched
$^{76}Ge$ experiment by the Heidelberg-Moscow collaboration
\cite{klapdor}. Interpreted in terms of a Majorana mass of the
neutrino, this implies $m_{ee}$ between 0.11 eV to 0.56 eV. If
confirmed, this result is of fundamental significance. For more
discussions of this result, we refer the reader to the report of the
double beta decay working group.

 A very different way to get
information on the absolute scale of neutrino masses is to study the spectrum of
the cosmic microwave background radiation (CMB), as well as to study the large scale
structure of the universe. This is discussed in the cosmology working group
report. Observations of CMB anisotropy
and surveys of large scale structure have set a limit
on the sum of neutrino masses $\sum m_i \leq 0.7-2$~eV 
\cite{hannestad}. More recent results from the Sloan Digital Sky
Survey (SDSS) place
the limit of $\sum m_i \leq 1.6$ eV. 
A point worth emphasizing is that the above result is valid for both
Majorana and Dirac neutrinos as long as the ``right-handed'' neutrinos
decouple before the BBN epoch and are not regenerated subsequently%
\footnote{ In the Dirac case the ``right-handed''
degrees of freedom are decoupled because of the smallness of the corresponding 
Yukawa couplings.}.

These limits already provide nontrivial information about neutrino masses: 
the limit $\sum_i m_{i}=0.7$~eV, if  taken at face value,
implies that each individual neutrino mass is smaller than $0.23$~eV, which 
is similar to the projected sensitivity of the 
proposed KATRIN experiment. 

It is clear from the above discussion that there are three urgent pieces
of information needed to answer the question of whether the neutrinos
are Majorana or Dirac fermions. Three experiments that play a crucial role in
this are (i) neutrinoless double beta decay search; (ii) determination of
the sign of $\Delta m^2_{13}$ and (iii) direct search for neutrino mass in
tritium decay (e.g.\ KATRIN) or similar decay experiments. In Table
\ref{tab:NatureOfNeutrinos}, we
present conclusions about the nature of neutrinos for different
outcomes of these three types of experiments (for KATRIN a
goal of 0.2 eV and for $\beta\beta_{0\nu}$ decay a goal of about 2 meV is
taken and $\Delta m^2_{13}= m^2_3-m^2_1$).

\begin{table}
\centering
\caption{Different possible conclusions regarding the
nature of the neutrinos and their mass hierarchy from the three
complementary experiments.}
\label{tab:NatureOfNeutrinos}
\begin{tabular}{|c||c||c||c|}
\hline
$\beta\beta_{0\nu}$ & $\Delta m^2_{13}$ & KATRIN & Conclusion \\ \hline
yes & $>0$ & yes & Degenerate, Majorana \\
yes & $>0$ & No & Degenerate, Majorana\\
 & & & or normal, Majorana with heavy particle contribution\\
yes & $<0$ & no & Inverted, Majorana \\
yes & $<0$ & yes & Degenerate, Majorana\\
no & $>0$ & no & Normal, Dirac or Majorana\\
no & $<0$ & no & Dirac\\
no & $<0$ & yes & Dirac \\
no & $>0$ & yes & Dirac \\ \hline
\end{tabular}
\end{table}

It is clear from Eq.~(\ref{V}) that for Majorana neutrinos,
there are three CP phases that characterize neutrino mixings and our 
understanding of the leptonic sector will remain incomplete without
knowledge of these \cite{boris,boris_2}.  There are two possible ways
to explore CP phases: (i) one way is to perform long baseline
oscillation experiments and look for differences between neutrino and
anti-neutrino survival probabilities \cite{minakata}; (ii) another way
is to use possible connections with cosmology. It has often been
argued that neutrinoless double beta decay may also provide an
alternative way to explore CP violation.

\subsection{Sterile neutrinos}

A question of great importance in neutrino physics is the number of
neutrino species. Measurement of the invisible $Z$-width in LEP-SLC
experiments tell us that only three types of neutrinos couple to the $W$
and $Z$ boson. They correspond to the three known neutrinos
$\nu_{e,\mu,\tau}$. This implies that if there are other neutrino
species, then they must have little or no interaction with the $W$ and
$Z$. They are called sterile neutrinos. So the question is: are there
any sterile neutrinos and, if so, how many?

In the Los Alamos Liquid Scintillation Detector (LSND) experiment,
neutrino oscillations both from a stopped muon (DAR) as well as the
one accompanying the muon in pion decay have apparently been observed.
The evidence from the DAR is statistically more significant and is an
oscillation from $\bar{\nu}_\mu$ to $\bar{\nu}_e$. The mass and mixing
parameter range that fits data is $ \Delta m^2 \simeq 0.2 -
2~\mathrm{eV}^2$, $\sin^22\theta \simeq 0.003-0.03$.  There are points
at higher masses specifically at 6 eV$^2$ which are also allowed by
the present LSND data for small mixings. The KARMEN experiment at the
Rutherford laboratory has very strongly constrained the allowed
parameter range of the LSND data. Currently the MiniBooNE experiment
at Fermilab is under way to probe the LSND parameter region.

Since this $\Delta m^2_\mathrm{LSND}$ is much larger than $\Delta
m^2_{12,23}$, the simplest way to explain these results is to add
one~\cite{caldwell,other} or two~\cite{sorel} sterile neutrinos. The sterile
neutrinos raise important issues of consistency with cosmology as well as
physics beyond the simple three neutrino picture and will be discussed in a
subsequent section.     

\subsection{Neutrino magnetic moment and neutrino decay}

A massive neutrino can have a magnetic moment. The presence of a
magnetic moment allows for new electromagnetic interactions between
neutrinos and other fermions of the Standard Model. In particular in
neutrino-electron scattering, in addition to the usual weak
interaction contribution, there will be a photon exchange contribution
to the scattering cross section.  The existing neutrino scattering
measurements therefore provide an upper limit on the neutrino magnetic
moment: $\mu_{\nu_e}\leq (1-1.3)\times 10^{-10}\mu_B$ where
$\mu_B~=~\frac{e}{2m_e}$ is a Bohr magneton. As we discuss in detail
later on, the magnetic moment is a sensitive measure of any new TeV
scale physics, i.e.\ if all physics beyond the Standard Model is at the
scale of grand unification or higher, the neutrino magnetic moments
will be of order $10^{-19}\mu_B\left(\frac{m_\nu}{1~eV}\right)$. Thus
any magnetic moment above this value implies the existence of new
physics at the TeV scale. A high precision search for a magnetic
moment is therefore very important for learning about physics just
beyond the Standard Model scale.

Neutrino magnetic moment also leads to new processes that can alter
our understanding of energy balance in astrophysical systems such as
in stars and supernovae~\cite{raffelt}. It can also affect
considerations involving the neutrinos in the early universe such as
the BBN. In sec. (\ref{mm}) we discuss more details on magnetic moment
and what one can learn from various proposed experiments.

The existence of a neutrino magnetic moment is also related to
neutrino decays. For instance if there is a cross-generational
structure to magnetic moment as will necessarily be the case if
neutrinos are Majorana fermions, then heavier neutrino species can
decay radiatively to the lighter ones. Such decays can be detectable
in astrophysical experiments. Present upper limits coupled with the
general idea about spectra of neutrinos from oscillation experiments,
imply that lifetimes of active neutrinos are larger than $10^{20}$
sec., much longer than the age of the universe. Such decays do not
therefore affect the evolution of the universe.

It is however possible that there are other scalar particles to which
the neutrinos decay; one such example is the majoron, which is a
Goldstone boson corresponding to the spontaneous breaking of a global
$B-L$ symmetry~\cite{cmp}. The decay to these scalar bosons may occur at
a faster rate than that to photons and may therefore have
astrophysical and cosmological implications~\cite{beacom}.

\section{The Questions}

The existing data on neutrinos have already raised very important questions,
such as the very different mixing angles, that are blazing new trails in
physics beyond that Standard Model. They are also helping to
 define sharp questions to be
addressed by near future experiments:
\begin{quote}
$\bullet$ Are neutrinos Dirac or Majorana?\\
$\bullet$ What is the absolute mass scale of neutrinos?\\
$\bullet$ How small is $\theta_{13}$?\\
$\bullet$ How ``maximal'' is $\theta_{23}$?\\
$\bullet$ Is there CP Violation in the neutrino sector?\\
$\bullet$ Is the mass hierarchy inverted or normal?\\
$\bullet$ Is the LSND evidence for oscillation true?  Are there sterile
      neutrino(s)?  
\end{quote}

In the near future, we hope to significantly improve the determination
of the elements of the neutrino mass-matrix, although some uncertainty
will still remain \cite{reconstruction}. Through neutrino oscillation
experiments, all three mixing angles $\theta_{12}, \theta_{23}$, and
$\theta_{13}$ are expected to be determined with good precision (this
is one of the main goals of next-generation neutrino oscillation
experiments), while there is hope that the ``Dirac phase" $\delta$ can
be probed via long-baseline $\nu_{\mu}\to\nu_e$ oscillation searches.
Neutrino oscillation experiments will also determine with good
precision the neutrino mass-squared differences ($\Delta m^2_{12}$ at
the 5\%--10\% level, $\Delta m^2_{13}$ [including the sign] at the few
percent level). In order to complete the picture, three other
quantities must also be measured, none of which is directly related to
neutrino oscillations.

One is the overall scale for neutrino masses. As already briefly
discussed, this will be probed, according to our current understanding,
by studies of the end-point spectrum of beta-decay, searches for
neutrinoless double beta decay, and cosmological observations (especially
studies of large-scale structure formation). Note that neutrinoless
double-beta decay experiments are sensitive to $|m_{\nu}^{ee}|$, 
i.e., they directly measure the absolute value of an element of $m_{\nu}$. 
The other two remaining observables are the
``Majorana" phases. 

Neutrinoless double beta decay experiments are
sensitive to a particular combination of masses, mixings and phases:
\begin{equation}
\left|m_{\nu}^{ee} \right|= \left|\cos^2\theta_{13} 
\left(|m_1|\cos^2\theta_{12}+|m_2|e^{-2i\phi_1}\sin^2\theta_{12}\right) 
+\sin^2\theta_{13}|m_3|e^{-2i(\phi_2-\delta)}\right|.
\end{equation}
In practice, however, it seems at least very challenging \cite{noMaj} to
obtain any information regarding Majorana phases from neutrinoless
double-beta decay, in part due to the fact that the relevant nuclear
matrix elements need to be computed with far more precision than has been
currently achieved. 

It must of course be made clear that
neutrinoless double-beta decay rate is related to the Majorana phases and
neutrino masses only under the assumption that the neutrino masses
are the only source of lepton-number violation. Second, only a
combination of the two
independent Majorana phases can be determined in this way. It is fair to
say that there is no realistic measurement one can look forward to making
in the near future that will add any information and help us disentangle
the ``other" Majorana phase. Third, it is curious to note that the effect
the Majorana phases have on the rate for neutrinoless double-beta decay
is CP-even. While Majorana phases can mediate CP violating
phenomena \cite{boris_2}, it seems unlikely that any of them can be
realistically studied experimentally in the foreseeable future.

In spite of all the uncertainty due to our inability to measure Majorana
phases, it is fair to say that we expect to correctly reconstruct several
features of the neutrino mass matrix \cite{reconstruction}, especially if
the overall mass-scale and the neutrino mass hierarchy are determined
experimentally. This will help us uncover whether there are new
fundamental organizing principles responsible for explaining in a more
satisfying way the values of the neutrino masses and the leptonic mixing
angles e.g. whether there are flavor (or family) symmetries, capable of
dynamically distinguishing the different generations of quarks and
leptons and/or whether there is quark-lepton unification at short
distances etc.

Answers from experiments will have crucial impact on the development
of the new Standard Model that incorporates newly discovered neutrino
mass.  Moreover, there are many deep theoretical questions that will
be influenced by data.  For example,
\begin{quote}
$\bullet$ We find that the value of $\theta_{13}$ is a good discriminator
  of models.\\
$\bullet$ Testing the seesaw hypothesis and discriminating between different
types of seesaw using lepton flavor violation.\\
$\bullet$ Are there new exotic interactions, possibly flavor-changing, for
  neutrinos?\\
$\bullet$ What are the admixtures of sterile neutrinos both heavy and
  light?\\
$\bullet$ Do neutrinos have magnetic moments?\\
$\bullet$ How can we understand the neutrinos' role in the origin of the
  cosmic baryon asymmetry?\\
$\bullet$ Can we use neutrinos as probes of other physics beyond the
  Standard Model?
\end{quote}
This document briefly addresses many of these questions.

\section{Neutrino Mass Models} 

To discuss neutrino masses, we have to specify if they are of Dirac or
Majorana type.

Dirac neutrinos require the existence of new right-handed
neutrinos that have not been observed.  The neutrino mass is generated
by the Yukawa coupling of left- and right-handed neutrinos to the
Higgs boson, in exactly the same fashion as the charged-lepton and
quark masses.  Even though this is a simple extension of the Standard
Model, the extreme smallness of the neutrino masses requires the Yukawa
couplings to be  of $O(10^{-13})$ or less, which most theorists believe
requires an explanation.  Some models of extra dimensions or
supersymmetry may offer an explanation as discussed later in this
report.

\begin{figure}[tbp]
  \centering
  \includegraphics[width=\textwidth]{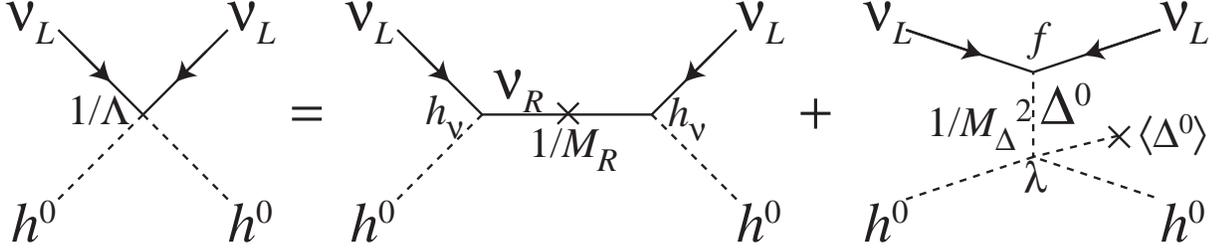}
  \caption{Seesaw mechanism that explains small Majorana neutrino mass
    by the exchange of GUT-scale particles.}
  \label{fig:seesaw}
\end{figure}

The Majorana mass, on the other hand, does not require new light
degrees of freedom, but rather a new higher-dimensional operator
Eq.~(\ref{eq:D5}) $\frac{1}{\Lambda} (LH)(LH)$.  The atmospheric
neutrino data require that $\Lambda \lesssim 10^{16}$~GeV, much lower
than the Planck or string scales.  It mostly likely means that there
are new heavy degrees of freedom whose interaction result in this
operator.  There are two such possibilities.  One is to exchange heavy
Majorana right-handed neutrinos that have (not small) Yukawa coupling
to the left-handed lepton and Higgs.  This mechanism is called the
seesaw mechanism \cite{Yanagida:1980} because the heavier right-handed
neutrino mass implies a lighter left-handed neutrino mass.  Another is
to exchange an $SU(2)_L$ triplet scalar coupled to $LL$ and $HH$,
possible in some SO(10) grand-unified models.  The latter possibility
is called type II seesaw mechanism.

In this section, we discuss models of Majorana neutrino masses and
mixings and their predictions on important quantities such as
$\theta_{13}$ and neutrinoless double beta decay rates.
 
\subsection{Bottom-Up Models}

Here, we discuss phenomenological models inspired by the data that in
turn give predictions to parameters not measured so far.

In Table I, we identify several textures for the neutrino mass matrix
that lead to the currently observed mass-squared differences and
mixing angles, and some of the measurements that will allow us to
identify which textures best describe nature. One caveat to the
usefulness of this approach is that we have made a choice of weak
basis where the charged lepton masses are diagonal and fundamental
theories need not manifest themselves in this basis. Nonetheless, by
studying some of these textures we can determine some of the
measurements (and how precise they should be) that will shed a
significant amount of light on the issue of interpreting neutrino
masses and mixing angles.

\renewcommand\arraystretch{0.9}
\begin{table}[tbp]
\caption{Different leading-order neutrino mass-textures and their
``predictions" for various observables. The fifth column indicates the
``prediction" for $|\cos2\theta_{23}|$ when there is no symmetry relating
the different order one entries of the leading-order texture (`n.s.'
stands for `no structure', meaning that the entries of the matrices in
the second column should all be multiplied by and order one coefficient),
while the sixth column indicates the ``prediction" for
$|\cos2\theta_{23}|$ \cite{theta_23_andre} when the coefficients of the leading order texture
are indeed related as prescribed by the matrix contained in the second
column. }
\label{texturetable}
\begin{tabular}{|c|c|c|c|c|c|c|} \hline
Case & Texture & Hierarchy  & $|U_{e3}|$ & $|\cos2\theta_{23}|$ (n.s.) &
$|\cos2\theta_{23}|$ & Solar Angle \\ \hline  
A & $\frac{\sqrt{\Delta m^2_{13}}}{2} \left(\begin{array}{ccc}0&0&0\\
0&1&1 \\ 0&1&1\end{array}\right)$& Normal &$\sqrt{\frac{\Delta
m^2_{12}}{\Delta m^2_{13}}}$ & O(1) & $\sqrt{\frac{\Delta
m^2_{12}}{\Delta m^2_{13}}}$ & O(1) \\ \hline
B &  $\sqrt{\Delta m^2_{13}}\left(\begin{array}{ccc}1&0&0\\
0&\frac{1}{2}&-\frac{1}{2} \\
0&-\frac{1}{2}&\frac{1}{2}\end{array}\right)$& Inverted & $\frac{\Delta
m^2_{12}}{|\Delta m^2_{13}|}$ & -- & $\frac{\Delta m^2_{12}}{|\Delta
m^2_{13}|}$ & O(1) \\ \hline
C &  $\frac{\sqrt{\Delta m^2_{13}}}{\sqrt{2}}
\left(\begin{array}{ccc}0&1&1\\ 1&0&0\\ 1&0&0\end{array}\right)$ &
Inverted & $\frac{\Delta m^2_{12}}{|\Delta m^2_{13}|}$ & O(1) &
$\frac{\Delta m^2_{12}}{|\Delta m^2_{13}|}$ & 
$\displaystyle \begin{array}{ll}
|\cos2\theta_{12}| \\
\sim\frac{\Delta m^2_{12}}{|\Delta m^2_{13}|}\end{array}$ \\ \hline
Anarchy & $\sqrt{\Delta m^2_{13}}\left(\begin{array}{ccc}1&1&1\\ 1&1&1 \\
1&1&1\end{array}\right)$ & Normal\footnote{One may argue that the
anarchical texture prefers but does not require a normal mass hierarchy.}
& $>0.1$ & O(1) & -- & O(1) \\ \hline
\end{tabular}
\end{table}

As is clear from Table~\ref{texturetable}, there are several completely different textures that
explain the current neutrino data. They differ, however, on the prediction for yet unknown
parameters. In particular, one can identify that knowledge of  the mass hierarchy and measurements of  
whether $|U_{e3}|^2\gtrsim0.01$ and/or $|\cos2\theta_{23}|\gtrsim0.01$
will allow us
 to determine the best path to follow as far as understanding neutrino masses and
 leptonic mixing is concerned.


\begin{table}[tbp]
  \caption{ \label{tabmeff1} The maximal values of  
    $\langle m\rangle_{\it eff}${} (in units of meV)
    for the NH  and IH spectra, and the minimal values of 
    $\langle m\rangle_{\it eff}$ (in units of  meV) for the IH and QD spectra,
    for the best fit values of the oscillation parameters and 
    $\sin^2\theta_{13} = 0.0$, $0.02$ and 
    $0.04$.
    The results for the NH and IH spectra 
    are obtained for 
    $|\Delta m^2_{23}| = 2.6 \times 10^{-3} \ eV^2~
    (2.0 \times 10^{-3} \ eV^2 -~{\rm values~in~brackets})$ and
    $m_1 = 10^{-4}$ eV, while
    those for the QD spectrum correspond to 
    $m_0 = 0.2$ eV. (From ref.~\protect\cite{PPaddendum}).
  }
\begin{center}
\begin{tabular}{|c|c|c|c|c|} 
\hline
\rule{0pt}{0.5cm} $\sin^2 \theta_{13}$
& ${\langle m\rangle_{\it eff}}_{\rm max}^{\rm NH}$ & 
${\langle m\rangle_{\it eff}}_{\rm min}^{\rm IH}$ &
${\langle m\rangle_{\it eff}}_{\rm max}^{\rm IH}$ &
${\langle m\rangle_{\it eff}}_{\rm min}^{\rm QD} $  
\\ \hline \hline
0.0   & 2.6 (2.6)   &  19.9 (17.3)    &  50.5 (44.2)   & 79.9 \\ \hline
 0.02 & 3.6 (3.5)   &  19.5 (17.0)    &  49.5 (43.3)   & 74.2 \\ \hline
 0.04 & 4.6 (4.3)   &  19.1 (16.6)    &  48.5 (42.4)   & 68.5 \\ \hline
\end{tabular}
\end{center}
\end{table}

As already noted, Majorana nature of massive neutrinos lead to
neutrinoless double beta decay processes of type $(A,Z) \rightarrow
(A,Z+2) + e^- + e^-$. This is subject of another working group report.
Therefore we do not discuss it in depth here except to note the very
interesting prediction that the currently contemplated precision of
the next generation of experiments will throw important light on the
different pattern of masses which is yet to be experimentally
determined. Given the current information on the neutrino oscillation
parameters, we summarize in Table \ref{tabmeff1} some typical
predictions for $\langle m\rangle_{\it eff}$ in different models.
Especially interesting is the lower bound on $\langle m\rangle_{\it eff}$ 
for the case of inverted hierarchy, a value that the next
generation experiments are supposed to be able to probe. 



\subsection{Grand Unified Models} 

One of the major ideas for physics beyond the Standard Model is
supersymmetric grand unification (SUSY GUT). It is
stimulated by a number of observations that are in accord with the
general expectations from SUSY GUTs : (i) A solution to the gauge
hierarchy problem i.e why $v_{wk}\ll M_{P\ell}$; (ii) unification of
electroweak, i.e.\ $SU(2)_L\times U(1)_Y$ and strong $SU(3)_c$ gauge
couplings assuming supersymmetry breaking masses are in the TeV range,
as would be required by the solution to the gauge hierarchy; (iii) a
natural way to understand the origin of electroweak symmetry breaking.

The unification of gauge couplings points to a unification scale around
$10^{16}$ GeV and simple seesaw intuition leads to a seesaw scale under
$10^{16}$ GeV in order to fit atmospheric neutrino data. This suggests that
the seesaw scale could be the GUT scale itself; thus the smallness of
neutrino mass could go quite well with the idea of supersymmetric
grand unification (although one can also get light neutrinos in, 
e.g., a model with a TeV  scale seesaw \cite{shrock}.). However, in
contrast with the items (i) through
(iii) listed above, the abundance of information for neutrinos makes
it a highly nontrivial exercise to see whether the neutrino mixings
indeed fit well into SUSY GUTs.  In fact, most GUT models proposed
before 1998 have been ruled out by the discovery of large mixing angles.
In turn, the freedom in constructing realistic GUT models allows many
different ways to explain current neutrino observations. Thus, even
though neutrino masses are solid evidence for physics beyond the
Standard Model, the true nature of this physics still remains obscure.
The hope is that the next round of the experiments will help to narrow
the field of candidate theories a great deal.

\begin{table}
\caption{The table lists some typical predictions for
$\theta_{13}$ in different SO(10) models and shows how the next
generation of experiments can narrow the field of possible
SO(10) unification models.}

\begin{tabular}{|c||c|}\hline
{\it {\bf 126} based models} & $\theta_{13}$ \\ \hline
Goh, Mohapatra, Ng & 0.18 \\
Chen, Mahanthappa & 0.15 \\ \hline
\end{tabular}
\begin{tabular}{|c||c|}\hline
{\it {\bf 16} based models} & $\theta_{13}$ \\ \hline
Albright, Barr & 0.014 \\Ross, Velasco-Sevilla & 0.07\\
Blazek, Raby, Tobe & 0.05 \\
\hline
\end{tabular}

\end{table}

While the $SU(5)$ group is enough to unify $SU(3)_c \times SU(2)_L
\times U(1)_Y$ in to a simple group, it does not unify the matter
content.  They are split into ${\bf 5}^* (d^c, L)$ and ${\bf 10}(Q,
u^c, e^c)$.  The right-handed neutrinos can be introduced to the model
but are not required.  On the other hand, the $SO(10)$ group unifies
the matter content into a single ${\bf 16}$, which in turn requires
right-handed neutrinos.  Their mass is naturally of the order of the
grand-unification scale, once $B-L$ is broken by a Higgs in either ${\bf
16}$
or ${\bf 126}$ representation, and their exchange produces the
neutrino mass operator Eq.~(\ref{eq:D5}).  The Higgs in ${\bf 126}$,
however, also contains a weak-triplet whose exchange can give rise to
the type-II seesaw as well.  We give a very small sample of the
different predictions for $\theta_{13}$ in models with either {\bf 16}
or {\bf 126} in Table IV and  a very incomplete list of references
in Ref.~\cite{so10,so1016}.

Here, we briefly summarize generic consequences of SO(10) models.  (i)
The neutrino mass hierarchy is normal although with type II
seesaw, the spectrum can be degenerate. In fact any evidence for a
degenerate neutrino spectrum would be an indication for type II seesaw
in general. (ii) They make definite
predictions about the mixing angle $\theta_{13}$ as given in Table IV
and often for the other mixing angles.  (iii) Neutrinos are Majorana.
The first two predictions can be tested by long-baseline neutrino
oscillation experiments, while the third by neutrinoless double beta
decay experiments.

\subsection{Renormalization Effects} 

In the study of these top-down models, the renormalization group (RG) evolution
may affect the neutrino masses and mixings significantly. Formalisms have been
developed to study it model-independently.  The RG equation of
the effective neutrino mass operator in the SM and MSSM
\cite{Chankowski:1993tx} can be translated  into differential equations for the
energy dependence of the mass eigenvalues, mixing angles and CP phases
\cite{Casas:1999tg}.  In the SM and in the MSSM with
small $\tan\beta$, the RG evolution of the mixing angles is negligible due to
the smallness of the $\tau$ Yukawa coupling. It is the stronger the more degenerate the mass spectrum is. For a strong
normal mass hierarchy, it is negligible even in the MSSM with a large
$\tan\beta$, but for an inverted hierarchy a significant running is possible
even if the lightest neutrino is massless. Furthermore, non-zero phases tend to
damp the running.  Typically, $\theta_{12}$ undergoes the strongest RG
evolution because the solar mass squared difference is much smaller than the
atmospheric one.  The RG equations for the CP phases show that whenever the
mixings run sizably, the same happens for the phases.

Apart from modifying the predictions of top-down models, RG effects also
open up new possibilities for model building, such as the radiative
magnification of mixing angles~\cite{balaji}. If one restricts oneself to 
the running below the lowest seesaw scale, $M_1$, significant magnification can 
occur only if $m_i\geq 0.1$~eV (a value observable in
$\beta\beta_{0\nu}$ decay). With the running above $M_1$, magnification can
be achieved for less degenerate light neutrino spectra, too (see e.g.\
\cite{Antusch:2002fr}).
RG effects can also cause important changes of the
input parameters for calculations of high-energy processes relevant for
leptogenesis.  Furthermore, they induce deviations of $\theta_{13}$ from
zero and $\theta_{23}$ from the maximal angle that provide an additional
motivation for planned oscillation experiments.  Given the expected
accuracy of these measurements, even relatively small RG effects are
interesting in this context.

\section{Leptogenesis and low energy CP phase in seesaw models}%
\label{sec:Leptogenesis}

Understanding the origin of matter is one of the fundamental questions
of cosmology the answer to which is most likely going to come from
particle physics. The seesaw mechanism is at the heart of one particle
physics mechanism and we discuss what we can learn about neutrino
physics as well as the pattern of right handed neutrino masses from the
observed baryon asymmetry.
   
Three ingredients are required to generate the observed Baryon
Asymmetry of the Universe: baryon number violation, CP violation and
some out-of-thermal equilibrium dynamics.  The seesaw
model \cite{Yanagida:1980}, which was introduced to give small neutrino
masses, naturally satisfies these requirements, producing the baryon
asymmetry by ``leptogenesis'' \cite{FY}.  It is interesting to
investigate the relation between the requirements of successful
leptogenesis, and the observable neutrino masses and mixing matrix.
In particular, does the CP violation that could be observed in
neutrino oscillations bear any relation to leptogenesis?
  
The idea of leptogenesis is to use the lepton number violation of the
$N_i$ Majorana masses $M_i$, in conjunction with the $B+L$ violation contained
in the Standard Model, to generate the baryon asymmetry.  The most
cosmology-independent implementation is ``thermal leptogenesis''
\cite{FY,BP,apostolos,gian}.

\subsection{Thermal Leptogenesis\label{sec:thermal}}  

If the temperature $T_{RH}$ of the thermal bath after inflation is $
\gappeq M_{1}$, the lightest $N_i$, $N_1$, will be produced by scattering.
If $N_1$ subsequently decays out of equilibrium, a CP asymmetry
\begin{equation}
  \epsilon_1 = \frac{\Gamma(N_1 \rightarrow L H) - \Gamma(N_1
  \rightarrow \bar{L} H^*)}{\Gamma(N_1 \rightarrow L H) + \Gamma(N_1
  \rightarrow \bar{L} H^*)}
\end{equation}
in the decay produces a net asymmetry of Standard Model leptons. This
asymmetry is partially transformed into a baryon asymmetry by the
non-perturbative $B+L$ violation.  Thermal leptogenesis has been studied
in detail \cite{BP,apostolos,gian}; the baryon to entropy ratio
produced is
\begin{equation}  
  Y_B \simeq C \kappa   \frac{n}{s} \epsilon_1   
  \label{thlep}  ~~,
\end{equation}  
where $\kappa \leq 1$ is an efficiency factor to be discussed in a
moment, $n/s \sim 10^{-3}$ is the ratio of the $N_1$ equilibrium
number density to the entropy density, and $\epsilon_1$ is the CP
asymmetry in the $N_{1}$ decay.  $C \sim 1/3$ tells what fraction of
the produced lepton asymmetry is reprocessed into baryons by the $B+L$
violating processes.  $Y_B$ depends largely on three parameters: the
$N_{1}$ mass $M_{1}$, its decay rate $\Gamma_1$, and the CP
asymmetry $\epsilon_1$ in the decay.  The decay rate $\Gamma_j$ of
$N_{j}$ can be conveniently parameterized as $ \Gamma_j =
\frac{[h_\nu^\dagger h_\nu ]_{jj} M_j} {8 \pi } \equiv
\frac{\tilde{m}_j M_j^2}{8 \pi v_{wk}^2}~~, $ where $\tilde{m}_j$ is
often of order of the elements of the $\nu_L$ mass matrix, although it
is a rescaled $N_1$ decay rate.

Eq.\ (\ref{thlep}) can be of the order of the observed $Y_B \sim 3
\times 10^{-11} $ when the following conditions are satisfied:

(i) $M_{1}$ should be $ \lappeq T_{RH}$.\footnote{In the so-called 
  `strong washout' regime, $T_{RH}$ can be an order of magnitude smaller 
  than $M_1$ \cite{Buchmuller:2004nz}.}  This temperature is
  unknown, but bounded from above in certain scenarios.
  
(ii) The $N_{1}$ decay rate $\propto \tilde{m}_1 $ should sit in a
  certain range.  $\tilde{m}_1$ must be large enough to produce an
  approximately thermal number density of $N_{1}$s, and small enough
  that the $N_{1}$ lifetime is of order the age of the Universe at $T
  \sim M_{1}$ (the out of equilibrium decay condition).  These two
  constraints are encoded in the efficiency factor $\kappa$.

(iii) $\epsilon_1$ must be $\gappeq 10^{-6}$.

\begin{figure}[tbp]
  \centering
    \includegraphics[height=5cm,width=8cm]{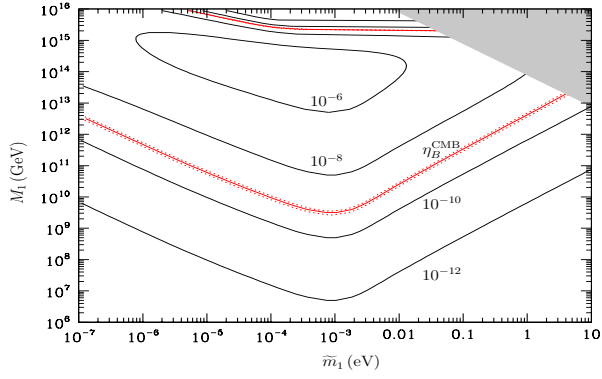} 
  \vspace*{-4mm}
  \caption{Contour plot of the baryon to photon ratio produced in 
    thermal leptogenesis in the plane of parameters $M_{1} $ and 
	$\tilde{m}_1$. The three (red) close-together correspond to the
    observed asymmetry as observed by WMAP. The plot is an updated version 
	of a plot of Ref.~\cite{di2}.}
  \label{YBBP}
\end{figure}

\begin{figure}[tbp]
  \centering
    \includegraphics[width=12cm]{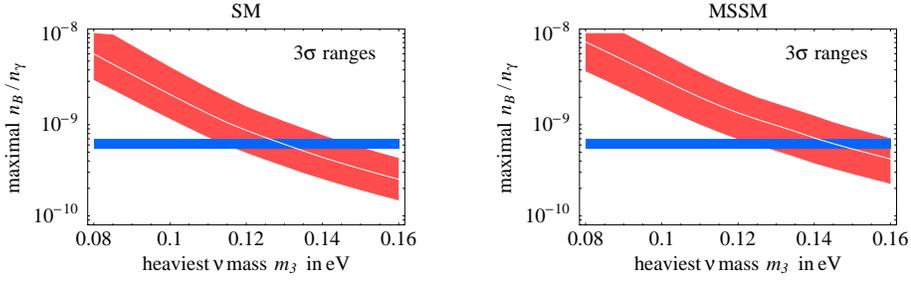} 
  \vspace{-2mm}
  \caption{Upper bound on the light neutrino mass scale, assuming
    hierarchical $M_i$, taken from \cite{strumia}.  The plot shows the
    measured baryon asymmetry (horizontal line) compared with the
    maximal leptogenesis value as function of the heaviest neutrino
    mass $m_3$.} 
  \label{m3strumia}
\end{figure}

The second requirement sets an upper bound on the mass scale of light
neutrinos.  The decay rate $\tilde{m}_1$ is usually $\sim m_2, m_3$;
for hierarchical light neutrinos, it naturally sits in the desired
range.  One can show that $m_1 \le \tilde{m}_1$, so $m_1
\lappeq 0.15 $ eV \cite{Fujii:2002jw,di2,gian,strumia} is required for thermal
leptogenesis in the type I seesaw.\footnote{Note that
Ref.~\cite{Buchmuller:2004nz} derives a somewhat tighter bound. Also
note that for type
II leptogenesis there is no longer any upper bound on $m_1$ (with
important implications for neutrinoless double beta decay).
Also in type II leptogenesis, the lower bound on $M_{1}$ could be
reduced by about an order of magnitude~\cite{antusch:2004xy}.}
This is shown in Fig.\ \ref{m3strumia}.
 
In the type I seesaw with hierarchical $N_i$, the third condition
imposes $M_{1} \gappeq 10^8$~GeV, because $\epsilon_1 \leq 3 M_{1}
(m_3 - m_1)/(8 \pi v_{wk}^2)$ in most of parameter space
\protect\cite{di2,strumia}.  If the $N_i$ are degenerate, with $\Delta
M_{ij} \sim \Gamma_i$, this bound on $M_{1}$ can be evaded
\protect\cite{apostolos,afsmirnov}.  For three $N_{i}$, the value of
$M_{1}$ has little implication on low energy neutrino observables.
If $\epsilon_1$ is maximal -- that is, $M_{1}$ close to its lower
bound, -- this sets one constraint on the 21 parameters of the type I
seesaw.  This has no observable consequences among Standard Model
particles, because at most 12 masses, angles and phases are
measurable, and $\epsilon$ can be maximized by choice of the nine
other parameters.  The situation is more promising \cite{dip} in SUSY
models with universal soft terms, where some of the 9 additional
parameters can contribute to slepton RGEs and thereby to charged lepton
flavour violating processes.

\subsection{Any relation with CP Violation in neutrino oscillations?}  
\label{phases}

The leptogenesis parameter $\epsilon_1$ is a $\CPV$ asymmetry,
suggesting a possible correlation with CP violation in $\nu$
oscillations (the phase $\delta$).  It turns out that there is no
linear connection between the MNSP phase and leptogenesis, i.e.\
leptogenesis can work when there is no $\CPV$ in MNSP, and measuring
low energy leptonic phases does not imply that there is CP violation
available for leptogenesis \cite{Branco:2001pq}.  In specific models, 
however, one may be able to relate the MNSP phase to the leptogenesis 
phase.

Turning to the type II seesaw case, the phase counting is same as in the
type I case and also there is in general no connection between low and
high energy CP violation either.  The number of CP phases can be
obtained by going to a basis in which both $m_\nu$ and $M$ are real
and diagonal since they are proportional to each other. Any CP
violation will then stem from the matrices $m_D$ and $m_\ell
m_\ell^\dagger$ (with $m_\ell$ being the charged lepton mass matrix).
Those two matrices posses in total 9 + 3 = 12 phases.  Since the type
II models have $M_\nu = v_L\ f_L$, with $f_L$ a symmetric $3 \times 3$
coupling matrix, which represents the coupling of weak-triplet
$\Delta_L$ Higgs field to leptons, there are new contributions to
leptogenesis.  The decay asymmetry in both $N_1$ and $\Delta$ decay
may arise from either $N_1$ or $\Delta$ exchange
\cite{O'Donnell:1994am,Hambye:2003ka}.
Thus, depending on which contribution dominates, four different
situations are possible \cite{Hambye:2003ka}.  If $M_1 \ll
M_{\Delta_L}$ and the conventional term ${\cal M_\nu}^I$ dominates
${\cal M_\nu}^{II}$, we recover the usual seesaw and leptogenesis
mechanisms and the statements given earlier apply.

\subsection{Resonant Leptogenesis}

 
If the mass difference between two heavy Majorana neutrinos happens to be much
smaller than their masses, the self-energy ($\varepsilon$-type) contribution to
the leptonic asymmetry becomes larger than the corresponding
($\varepsilon'$-type) contribution from vertex
effects~\cite{FPSCRV,apostolos}.  Resonant leptogenesis can occur when this
mass difference of two heavy Majorana neutrinos is of the order of their decay
widths, in which case the leptonic asymmetry could be even of order
one~\cite{apostolos,APTU}.
 As a result, one can maintain the RH neutrino
masses
around the GUT scale \cite{afsmirnov} or one can contemplate the possibility
that the heavy neutrino mass scale pertinent to thermal leptogenesis is
significantly lower being in the TeV~energies~\cite{apostolos}. This of course
requires a different realization of the seesaw mechanism \cite{mv} but it can
be in complete accordance with the current solar and atmospheric neutrino
data~\cite{APTU}.

The magnitude of the $\varepsilon$-type CP violation occurring in the
decay of a heavy Majorana neutrino $N_i$ is given by \cite{apostolos},
\begin{equation}
  \varepsilon_{N_i} = \frac{\mathrm{Im}
    (h^{\dagger}\,h_{\nu})^2_{ij}}{(h_{\nu\dagger}\,h_{\nu})_{ii}
    (h_\nu^{\dagger}\,h_{\nu})_{jj}}\,\frac{(M_{i}^2-M_{j}^2) M_{i}
    \Gamma^{(0)}_{j}}{(M_{i}^2-M_{j}^2)^2 + M_{i}^2
    \Gamma^{(0)\,2}_{j}}\,,
  \label{eps}
\end{equation}
where $\Gamma^{(0)}_{i}$ is the tree level total decay width of
$N_i$. It is apparent that the CP asymmetry will be enhanced, possibly
to $\varepsilon \sim 1$, provided if the first factor above is order
unity and if $M_{2}-M_{1} \sim
\frac{1}{2}\,\Gamma^{(0)}_{1,2}$.  It is important to note that
Eq.~(\ref{eps}) is only valid for the mixing of two heavy Majorana
neutrinos. Its generalization to the three neutrino mixing case is
more involved and is given in \cite{APTU}.

Successful leptogenesis requires conditions out of thermal
equilibrium.  To quantify this, we introduce the parameter, $K_i =
\Gamma^{(0)}_{i} / H(T=M_{i})$ where $H(T)$ is the Hubble
parameter.  $K_i$ should be smaller than a certain value, $K_i^{\rm
  max}$ for successful leptogenesis.  Using the parameter
$\widetilde{M}_i$ defined in Section \ref{sec:thermal}, it can be
re-expressed as $\widetilde{M}_i \lesssim 10^{-3}\, K^{\rm
  max}_i~{\rm eV}$.

Resonant leptogenesis can be successful with values of $K_i^{\rm max}$
larger than $1000$ \cite{APTU}.  This has
implications for leptogenesis bounds on the absolute mass scale of the
light neutrinos. If a large, $\gtrsim 0.2\,\mathrm{eV}$, Majorana mass
was seen in neutrinoless double beta decay, this could be naturally
accommodated with resonant leptogenesis.

The conditions for resonant leptogenesis can be met in several ways.
For instance, the `heavy' Majorana neutrinos can be as light as 1 TeV \cite{APTU}.
SO(10) models with a type III seesaw mechanism naturally predict pairs
of nearly degenerate heavy Majorana neutrinos suitable for resonant
leptogenesis \cite{mv,AB}.

Soft SUSY breaking terms can give small mass differences between
sneutrinos in soft leptogenesis \cite{DGRGKNR}. Resonant effects allow
sneutrino decay to generate the required CP asymmetry.  A model of
neutrino mass from SUSY breaking has also been shown to naturally lead
to conditions suitable for resonant leptogenesis \cite{THJMRSW}.

\subsection{Dirac Leptogenesis}\label{sec:DiracLeptogenesis}

In passing, we would like to mention that lepton number, or, more precisely,
$B-L$ has not necessarily to be violated in order to explain our existence,
i.e.\ the observed baryon asymmetry. In the context of neutrino-based
baryogenesis mechanisms, one can exploit the fact that only left-handed
particles couple to the sphalerons. It has been shown that, in the case of
Dirac neutrinos, lepton number can be stored in the right-handed neutrinos
during the washout \cite{Dick:1999je}. Thus, baryogenesis can work even if
$B-L$ is conserved. In particular, the requirement of successful baryogenesis
does not imply that neutrinos have to be Majorana particles.

\section{Issues beyond the minimal three neutrino picture}

\subsection{ Neutrino magnetic moments}\label{mm}

Once neutrinos are massive, they can have magnetic moments.  Magnetic
moment always connects one species of neutrino with another. When an
active neutrino ($\nu_{e,\mu,\tau}$ ) connects with an active
neutrino, we will call it a Majorana type magnetic moment. On the
other hand when one of the $\nu_i$ is a sterile neutrino, we will call
it Dirac moment. The two have fundamentally different physical
implications.

Neutrino magnetic moments can be directly measured in terrestrial
experiments using the neutrino beam from the Sun as in Super-K
\cite{superK} or with neutrinos from close by nuclear reactors as in
the MUNU \cite{munu} and in the Texono \cite{texono} experiments
because the presence of magnetic moment gives additional contribution
to neutrino scattering off electrons. These experiments have put upper
bounds of the order of $10^{-10} \mu_B$ on the effective neutrino
magnetic moment where $\mu_B$ is Bohr magneton (~$=\frac{e}{2m_e c}$).
It is also possible to put bounds on $\mu_{ij}$ from SNO-NC data
\cite{sno} using the fact that neutrinos with non-zero magnetic
moments can dissociate deuterium \cite{grifols} in addition to the
weak neutral currents. The bounds established from SNO-NC data do not
depend upon the oscillation parameters unlike in the case of Super-K.
However the bounds are poorer due to the large uncertainty in our
theoretical knowledge of the theoretical $^8B $ flux from the Sun
\cite{bahcall}.

In minimal extensions of the Standard Model that include neutrino
mass, the value of the neutrino magnetic moment is
$10^{-19}(m_\nu/{\rm 1 \ eV})\mu_B$\cite{shrock1}.
However new physics around a TeV tends to give larger values for the
magnetic moment \cite{shrock} and therefore search for magnetic moment
is a sensitive indicator of new physics near the TeV scale.  The
effective magnetic moment of the neutrinos can get substantially
enhanced in a certain class of extra dimensions models. Searching for
$\mu_{\nu}$ can therefore be used to put limits on theories with extra
dimensions. In particular, in a reactor experiment that searches for
differential cross section for $\nu_e-e$ scattering as a function of
the electron recoil energy, the extra dimension models produce a
distortion of the spectral shape, which can therefore be a crucial
signature of low fundamental scale, large extra dimension models.

\subsection{The search for {\em other} light neutrinos}

A neutrino that does not participate in Standard Model interactions
(sterile) might seem of little interest, but this concept includes
reasonable theoretical constructs such as right-handed neutrinos
themselves.  Furthermore, the hypothesis of `sterility' concerns the
weak forces; gravity is expected to be felt anyway, and we cannot
exclude that the `sterile' neutrino participates in new forces,
perhaps, mostly coupled to quarks; or carried by new heavy mediators;
or that sterile neutrinos have preferential couplings with new
particles -- say, with majorons.  Even putting aside these
possibilities, we can probe sterile neutrinos by the search for
observable effects due to their mixing with the ordinary neutrinos. In
this section, we will further restrict our attention on `light'
sterile neutrinos (say, below $10$~eV) and discuss the impact on
oscillations.  We make extensive reference to ref.~\cite{cmsv}, an
updated overview on the phenomenology of one extra sterile neutrino.

\begin{figure}[tbp]
\includegraphics[width=7cm]{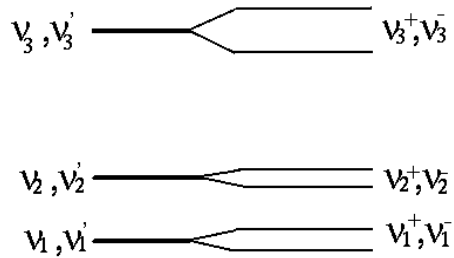}
\vspace{-3mm}
\caption{The double degeneracy between 
mass eigenstates of ordinary and mirror world 
($\nu_i$ and $\nu'_i$) is lifted when 
the small mixing terms are included in the $6\times 6$ mass matrix.
The new mass eigenstates ($\nu_i^+$ and $\nu_i^-$)
are in good approximation maximal superpositions  
of $\nu_i$ and $\nu'_i$. 
\label{fig:mirr}} 
\end{figure}

Many extensions of the Standard Model incorporate particles behaving
as sterile neutrinos.  The main question is \cite{pl} why these are
light.  Models with mirror matter (and mirror neutrinos) offer a
straightforward answer: ordinary and mirror neutrinos are light for
the same reason.  It is easy to arrange a `communication' term between
ordinary and mirror worlds, e.g., due to the operator $\sim
\nu\phi\nu'\phi'/M_{\rm Planck}$.  This leads to long-wavelength
oscillations into sterile neutrinos (see Fig.\ \ref{fig:mirr}, from
\cite{bnv}).  There are many other possibilities.  Already with mirror
matter, the VEV $\langle \phi'\rangle$ could be different from
$\langle \phi\rangle=174$~GeV, and this has important consequences for
the phenomenology \cite{zr}.  Alternatively, one could guess on
dimensional grounds the value ${\rm TeV}^2/M_{\rm Planck}$ as the mass
(or mixing) of sterile neutrinos, and relate the TeV-value, e.g., to
supersymmetry breaking \cite{alyos}.  Understanding neutrino mass in
extra dimension models also requires the existence of light sterile
neutrinos.

In the following discussion of phenomenology, we will be concerned
mostly with oscillations.  However, the implications can be also
elsewhere.  To see that, it is sufficient to recall that when we add 3
sterile neutrinos we can form Dirac masses, which means that there is
no contribution to neutrinoless double beta decay process.

\paragraph{Terrestrial oscillation experiments}

Broadly speaking, there are two types of terrestrial experiments.  The
first one includes several disappearance experiments and LSND; the
second one includes atmospheric neutrinos and long baseline
experiments. The first type is sensitive mostly to the mixing of
$\nu_e$ and a sterile state, the other one also to $\nu_\mu$ or
$\nu_\tau$.  Both types of experiments probe only relatively large
mixing angles, $\theta_s\sim 0.1$.  Sterile neutrinos within the
sensitivity regions are disfavored if standard cosmology (mostly BBN)
applies; further important tests will be done by CMB+LSS or BBN data.
None of these experiments {\em alone} requires the existence of
sterile neutrinos.  A case for sterile neutrinos can be made
interpreting in terms of oscillations LSND together with solar and
atmospheric anomalies \cite{bgg}. The hypothesis that LSND signal is
due to a relatively heavy and mostly sterile neutrino should be
regarded as conservative \cite{ales}, even though it leads to some
problems with disappearance in terrestrial experiments, and
interesting predictions for cosmology (BBN and CMB+LLS spectra). In
view of this situation, the test of the LSND result is of essential
importance. At the same time, we should not forget that sterile
neutrinos could manifest themselves in other manners.

\paragraph{Solar and KamLAND neutrinos}
The solar and KamLAND data can be explained 
well without sterile neutrinos. Even more, the `LMA' solution
received significant confirmations: the sub-MeV
energy regions  have been probed by Gallium experiments and the 
super-MeV ones by SNO and Kamiokande, and LMA is in 
agreement with KamLAND. Thus we are led to consider  
minor admixtures of sterile neutrinos, presumably not more than 20~\%.
In many interesting cases sterile neutrinos are invisible
at KamLAND but affect the survival probability of solar neutrinos.
Quite 
generally, to test the hypothesis of oscillations into sterile 
states it would be important to improve on (or measure precisely) 
the fluxes from Beryllium and pp-neutrinos.



\paragraph{Ultra-high energy neutrinos}
Although there is a great deal of interest in the search for ultra-high energy
neutrinos, the number of reasonable (or even, less reasonable) mechanisms 
that have been discussed to produce them 
is not large. The reason 
is that neutrinos are 
produced along with electromagnetic radiation, that can be observed 
in a variety of ways, even when this is reprocessed. 
Following this line of thought, the 
astrophysical mechanism that can be conceived to overcome 
such a stricture is the concept of a `hidden source'. 
Another escape from this constraint involves sterile neutrinos.
Indeed, if there are ultra-high energy 
mirror neutrinos, they inevitably 
oscillate into neutrinos from our world on 
cosmic scales \cite{bv}. This scenario can provide 
intense fluxes of ultra-high energy 
neutrinos, subject only to the observable electromagnetic radiation from 
their interaction with the relic neutrino sea.

\subsection{Supersymmetry and neutrinos:}

Neutrino masses are not the only motivation to extend the Standard
Model. One also likes to extend it in order to solve the gauge hierarchy
problem. Models of low-energy supersymmetry are
attractive candidates for the theory of TeV scale physics.
In the minimal supersymmetric extension of the Standard Model
(MSSM) neutrinos are massless. Thus, we need to consider
supersymmetric extensions of the Standard Model that allow for
neutrino masses.

There are basically three questions we like to answer when we talk
about the relations between supersymmetry and neutrinos:

(i)
Can successful predictions for neutrino masses of non-supersymmetric
extensions of the Standard Model be retained once these models are
supersymmetrized? In particular, can supersymmetry help in making
such models more motivated?

(ii)
Are there models where neutrino masses arise only due to
supersymmetry?

(iii)
Are there interesting phenomena in the slepton sector that can shed
light on the issue of neutrino masses, lepton number violation and
lepton flavor violation?

In the following we briefly describe two frameworks where neutrino
masses are tightly connected to supersymmetry.  We also discuss two
effects, that of charged lepton flavor violation and sneutrino--antisneutrino 
oscillation, that can help us disentangle the origin of neutrino
masses using supersymmetric probes.

\subsubsection{Seesaw Mechanism and Charged Lepton Flavor Violation}

In the Standard Model, there is no lepton flavor violation.  Neutrino
oscillation experiments have revealed that flavour is much more
violated in the lepton than in the quark sector.  However if one
simply extends the Standard Model by the addition of an appropriate
neutrino mass matrix in a gauge invariant manner, the magnitude of charged
lepton flavor violation is very small (the branching ratio being given
by $\big(\frac{m_{\nu}}{m_W}\big)^4$). The situation remains
unchanged even when the seesaw mechanism is used to generate neutrino
masses, since the seesaw scale is very high. However, if the theory is
supersymmetric, the flavor mixings in either the Dirac neutrino mass
matrix or the RH neutrino mass matrix (or both) can transmit flavor
mixings to the slepton sector which can then lead to lepton flavor
violating processes such as $\mu\rightarrow e+\gamma$ and
$\tau\rightarrow \mu +\gamma$\cite{bm}. Current bounds for
$B(\mu\rightarrow
e+\gamma)\leq 1.2 \times 10^{-11}$ and $B(\tau\rightarrow
\mu+\gamma)\leq 2\times 10^{-7}$.  These limits are expected to be
pushed down to the level of $10^{-14}$ and $10^{-8}$ level. For
reasonable values for the supersymmetry parameters, seesaw models can
predict these branching ratios at these levels. 
Combined with supersymmetry searches at the LHC, one can hope to probe the
validity of the seesaw mechanism. One exception \cite{apo} to this is the
class of models with TeV scale seesaw \cite{mv}, where even without
supersymmetry, charged lepton flavor violation could be large.

\subsubsection{Neutrino masses from R-parity violation}

Neutrino masses from R-parity violation have been extensively studied.
Here we briefly summarize the main results \cite{Grossman:2003gq}.
Once R-parity is violated there is no conserved quantum number that
would distinguish between the down-type Higgs doublet and the lepton
doublets. Thus, these fields in general mix. Such mixing generates
neutrino masses; in fact, they generically produce too large masses.
One neutrino gets a tree level mass which depends on the mixings
between the Higgs and the sneutrinos. The other two neutrinos get
their masses at the one loop level, and thus their masses are smaller
by, roughly, a loop factor.  The most attractive feature of R-parity
violation models of neutrino masses is that they naturally generate
hierarchical neutrino masses with large mixing angles. This is due to
the fact that only one neutrino gets a mass at tree level, while the
other neutrinos only acquire loop induced masses. Numerically,
however, the predicted mass hierarchy is in general somewhat too
strong.  The biggest puzzle posed by R-parity violation models is to
understand the smallness of the neutrino masses. There must be a
mechanism that generates very small R-parity violating couplings.
There are several ideas of how to do it. For example, the small
R-parity violation couplings can be a result of an Abelian horizontal
symmetry \cite{Banks:1995by} or left-right SUSY \cite{kuchi}.

\subsubsection{Neutrino masses from supersymmetry breaking}

The smallness of neutrino masses can be directly related to the
mechanism of supersymmetry breaking, in particular to the mechanism
that ensures a weak scale $\mu$ parameter
\cite{Arkani-Hamed:2000bq,Arkani-Hamed:2000kj,Borzumati:2000ya}.  In
general, there is no reason why the MSSM $\mu$ parameter is of the
order of the weak scale. Generically, it is expected to be at the
cut-off scale of the theory, say the Plank or the GUT scale.
Phenomenologically, however, $\mu$ is required to be at the weak
scale. One explanation, which is known as the Giudice-Masiero
mechanism, is that a $\mu$ term in the superpotential is not allowed
by a global symmetry. The required effective weak scale $\mu$ is
generated due to supersymmetry breaking effects.

The Giudice-Masiero mechanism can be generalized to generate small
neutrino masses. It might be that the large Majorana mass term that
drives the seesaw mechanism is forbidden by a global symmetry.
Effective Majorana mass terms for the right handed neutrinos, of the
order of the weak scale, are generated due to supersymmetry breaking.
The same global symmetry can also suppress the Dirac mass between the
right and left handed neutrinos. Then, the left handed neutrinos have
very small Majorana or Dirac masses as desired.  The emerging neutrino
spectrum depends on the exact form of the global symmetry that is used
to implement the Giuduce-Masiero mechanism. Nevertheless, the feature
that the left-handed neutrino masses are very small is generic.

\subsubsection{Sneutrino oscillation}
Supersymmetric models can also lead to sneutrino--antisneutrino mixing and
oscillation \cite{Grossman:1997is}.  This phenomena is analogous to the
effect of a small $\Delta S=2$ perturbation to the leading $\Delta
S=0$ mass term in the $K$-system which results in a mass splitting
between the heavy and light neutral $K$ mesons. The very small mass
splitting can be measured by observing flavor oscillations.  
The sneutrino system can exhibit similar behavior. The lepton number
is tagged in sneutrino decay using the charge of the outgoing lepton.
The relevant scale is the sneutrino width. If the sneutrino mass
splitting is large, namely when
$x_{\tilde \nu} \equiv {\Delta m_{\tilde \nu} / \Gamma_{\tilde \nu}}
\gtrsim 1$,
and the sneutrino branching ratio into final states with a charged
lepton is significant, then a measurable same sign dilepton signal is
expected.  Any observation of such oscillation will be an evidence for
total lepton number violation, namely for Majorana neutrino masses.

\subsection{Neutrinos in extra dimensions}

The pioneering idea by Kaluza and Klein~(KK)~\cite{KK} that our world
may have more than four dimensions has attracted renewed interest over
the last ten years~\cite{ADD,DDG1}.  The possible existence of extra
dimensions has enriched dramatically our perspectives in searching for
physics beyond the Standard Model.  Obviously, extra dimensions have
to be sufficiently compact to explain why they have escaped detection
so far, although their allowed size is highly model-dependent.  This
means that the derived constraints not only depend on the number of
the fields sensitive to extra dimensions but also on the geometry
and/or the shape of the new dimensions.

Models with large extra dimensions generically have a low
fundamental scale, and it is often the case that the 
seesaw mechanism cannot be properly implemented. 
An alternative way to understand small neutrino masses is
to introduce singlet neutrinos that propagate  in  a higher $[1 +
(3+\delta)]$-dimensional space (where $\delta$ is the number
of the additional  spatial compact  dimensions).  In this  formulation,
the  ordinary SM particles  reside in  a $(1+3)$-dimensional Minkowski
subspace, which is called the wall.   The overlap  of  their
wave-functions with the bulk neutrinos  is suppressed by the volume of
the   extra-dimensional  space  $(R\,  M_F)^{\delta/2} \approx  M_{\rm
P}/M_{\rm  F}$,   where $R$  is  the   common compactification radius,
$M_{\rm F}$ is  the fundamental gravity  scale and  $M_{\rm P} \approx
10^{16}$~TeV is the usual Planck mass.  This volume-suppression factor
gives  rise to effective  neutrino Yukawa couplings that are naturally
very small, i.e.\ of  order $M_{\rm  F}/M_{\rm  P} \sim  10^{-15}$, for
$M_{\rm F} = 10$~TeV, although  the original higher-dimensional Yukawa
couplings  of the theory could   be of order  unity.

\begin{figure}[tbp]
  \centering
    \includegraphics[width=0.8\textwidth]{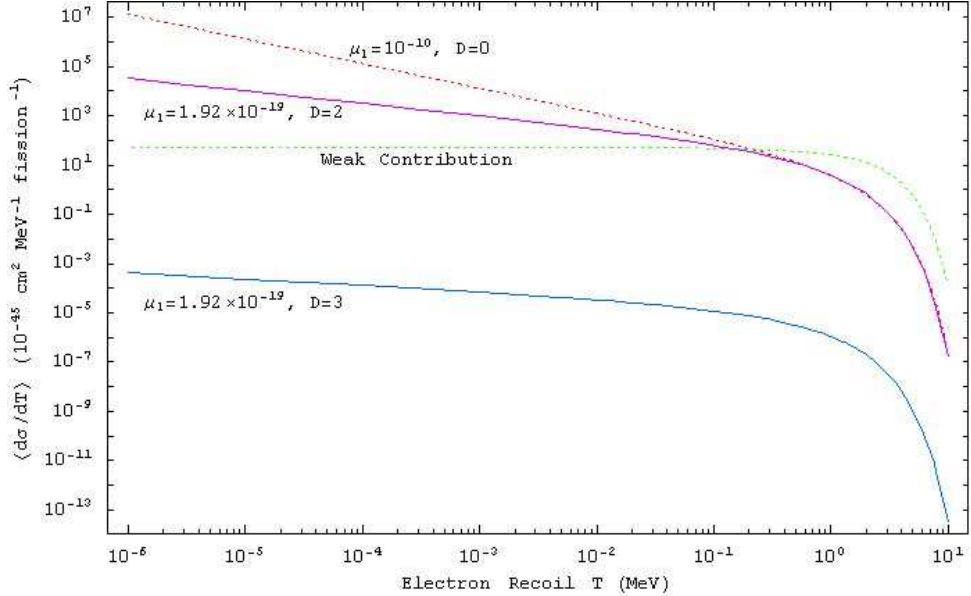}
    \vspace{-2mm}
    \caption{The figure shows the contribution of a neutrino magnetic
      moment for the case of single Dirac neutrino, for two large
      extra dimensions (and a comparison between the two) to
      differential cross section $\frac{d\sigma}{dT}$ (where $T$ is
      the electron recoil energy) for neutrino electron scattering and
      compares it to the case of one right handed neutrino
      (``Standard Model with one right handed neutrino").\label{fig:cstr1}}
\end{figure}

There are several generic consequences of these models:

 (i) There is a closely spaced tower of sterile neutrinos in
  such models which can be emitted in any process where the final
  state is a sterile neutrino. A typical example is the magnetic
  moment contribution to $\nu_e-e$ scattering in a
  reactor \cite{ng,ynm}. Reactor searches for magnetic moment can
  therefore shed light on the size of extra dimensions (see Fig.\ \ref{fig:cstr1}).

 (ii) When neutrinos travel
  through dense matter there can be MSW resonances \cite{DS} that can
  rise to a dip pattern \cite{DS,cmy} in the neutrino survival
  probability corresponding to energies spaced by $E\approx \Delta
  m^2_{\nu_F\nu_{KK}}/2\sqrt{2}G_F N_e$ (i.e.\ $E$, $4E$,
$9E$,...) since typically the survival probability goes like
  $e^{-c\frac{\Delta m^2}{E}}$. For solar
  neutrinos, such dip structure is quite pronounced \cite{cmy}.  In
  the hierarchical pattern for neutrino masses, this would correspond
  to $E\approx $ 10 MeV for densities comparable to solar core. The
  value of the energy clearly depends on the size of the extra
  dimensions; therefore looking at neutrinos of different energies
  such as those from Sun, atmosphere and distant galaxies, one can
  probe different sizes of the extra dimensions.

 (iii) The cumulative effect of the neutrino KK tower also leads to
enhanced flavor violating effects \cite{apo1}.

 (iii) One may add lepton-number violating bilinears of the
  Majorana type in the Lagrangian~\cite{DDG2}, e.g.~operators of the
  form $N^T C^{(5)-1} N$, where $C^{(5)} = - \gamma_1 \gamma_3$ is the
  charge conjugation operator, which can then add new contributions to
  neutrinoless double beta decay. These models provide other sources
  of both lepton flavor violation as well as lepton number violation
  that can be experimentally interesting.

\section{Exotic physics and neutrinos}

\subsection{ New long range forces}

The possible existence of new long range forces has always been an
interesting one in particle physics.  A special class of long range
forces which distinguish between leptonic flavors has far reaching
implications for neutrino oscillations \cite{am,grifols-masso}
which may be used as probes of such forces. Anomaly considerations
leave a limited choice for such forces, i.e., the ones coupling to
$L_i-L_j$ (where $i,j= e,\mu,\tau$). It is possible in this case to
have long range forces with range of the order of the Earth-Sun
distance.  Such forces would induce matter effects in terrestrial,
solar and atmospheric neutrino oscillations. For example, the
electrons inside the Sun generate a potential $V_{LR}$ at the earth
surface given by
\begin{equation} \label{vlr}
  V_{LR}=\alpha {N_e\over R_{es}}\approx (1.04 \times 10^{-11}
  eV)\left({\alpha\over 10^{-50}}\right) ~, 
\end{equation} 
where $\alpha\equiv {g^2\over 4 \pi}$ corresponds to the gauge
coupling of the $L_{e}-L_{\mu,\tau}$ symmetry , $N_e$ is the number of
electrons inside the Sun and $R_{es}$ is the Earth-Sun distance
$\approx 7.6 \times 10^{26} GeV^{-1}$. The present bound on the
$Z$-dependent force with range $\lambda\sim 10^{13}$ cm is given by
$\alpha< 3.3\times 10^{-50}$. Eq.\ (\ref{vlr}) then shows that the
potential $V_{LR}$ can introduce very significant matter-dependent
effects in spite of the very strong bound on $\alpha$. One can define
a parameter $\xi\equiv {2 E_\nu V_{LR}/ \Delta m^2}$ which measures
the effect of the long range force in any given neutrino oscillation
experiment. Given the terrestrial bound on $\alpha$, one sees that
$\xi$ is given by $\xi_{atm}\sim 27.4$ in atmospheric or typical long
baseline experiments while it is given by $\xi_{solar}\sim 7.6$ in the
case of the solar or KamLAND type of experiments. In either case, the
long range force would change the conventional oscillation analysis.
The relatively large value of $\alpha$ suppresses the oscillations of the
atmospheric neutrinos. The observed oscillations then can be used to
put stronger constraints on $\alpha$ which were analyzed in \cite{am}.
One finds the improved 90\% CL bound: $\alpha_{e\mu}\leq 5.5\times
10^{-52}$, $\alpha_{e\tau}\leq 6.4 \times 10^{-52}$,
in case of the $L_{e}-L_{\mu,\tau}$ symmetries respectively.

\begin{figure}[tbp]
  \centering
  \includegraphics[width=0.5\textwidth]{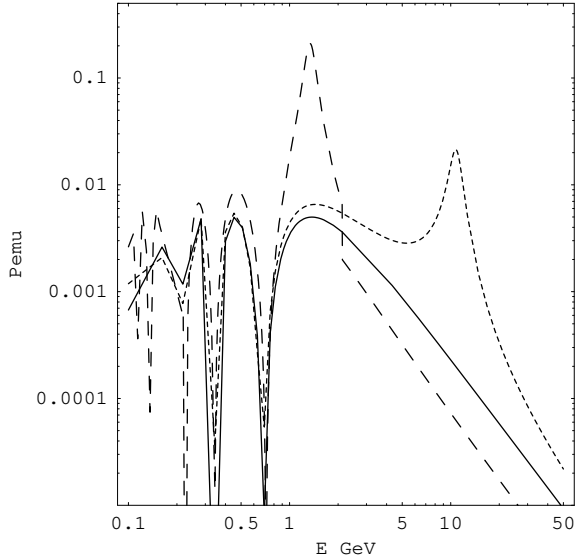}
  \vspace*{-7mm}
  \caption{{\small The long baseline neutrino oscillation probability
      $P_{e\mu}$ in case of vacuum (solid), the earth matter effects
      (dotted) and with inclusion of the long range potential $V_{LR}$
      (dashed). The plotted curves correspond to a baseline of 740 km,
      $\Delta m_{32}^2=2.5 \times 10^{-5} eV^2$, $\Delta m_{21}^2=7.0
      \times 10^{-5} eV^2$, ($\theta_{12},\theta_{23})=(32^0,45^0)$,
      $\alpha_{e\mu}=5.5 \times 10^{-52}$ and $\sin \theta_{13}=0.05$.
    }}
  \label{amfig1}
\end{figure}

Although these bounds represent considerable improvement over the
conventional fifth force bound, they still allow interesting effects
which can be used as a probe of such long range forces in future long
baseline experiments with super beam or at neutrino factories. As a
concrete example, the influence of the $L_e-L_\mu$ gauge interactions
on the long baseline oscillations of muon neutrinos of ${\cal O}$(GeV)
energy. The survival probability in this case as a function of energy
is given in Fig.\ \ref{amfig1}.

\subsection{Non-standard Neutrino Neutral Current Interactions}

The latest results of neutrino oscillation experiments indicate that
the conversion mechanism between different neutrino flavors is driven
by a non-vanishing mass difference between mass eigenstates together
with large mixing angles between families.  However, these conclusions
are achieved supposing that no non-standard neutrino interactions
(NSNI) are present. The inclusion of NSNI can modify the
characteristics of neutrino conversion, and in general large values of
NSNI parameters worsen the quality of the fit to data. We can then use
neutrino oscillation experiments to set limits to NSNI parameters.

The atmospheric neutrino data are well described by the oscillation
driven by one mass scale, $\Delta m^2_{32}$, and with maximal mixing
between second and third families.  The addition of sufficiently
  large NSNI on top of the masses and mixing can be expected to spoil
  the fit. A two-family ($\nu_\mu$, $\nu_\tau$)
  analysis~\cite{fornengo} constrains the flavor-violating NSNI to be
  less than a few percent of the standard weak interaction. A
  generalization to the three-family analysis~\cite{atm3fam}, however,
  significantly relaxes this bound and in fact allows NSNI comparable
  in strength to the standard model interactions.

The oscillation of solar neutrinos is driven by only one mass scale,
$\Delta m_{21}^2$. 
The upper limits on the flavor-conserving NSNI are at tens of percents
of the standard weak interaction and hence are surprisingly
weak~\cite{holanda2}. Moreover, the
  flavor-changing NSNI is likewise weakly constrained~\cite{pena1}.

Apart from phenomena that involve neutrino oscillations, bounds on
NSNI can also come from the bounds of such non-standard interactions
on the charged leptons. One should be careful in translating such
bounds to the neutrino sector, since one is only
be possible if details regarding the model that generates the
non-standard interactions are known. Recent analyses of such bounds can be found
in~\cite{bergmann1,holanda3,rossi,pena2}.

One can argue that here is a hint for NSNI. The NuTeV experiment \cite{Zeller:2001hh} at
Fermilab has measured the ratios of neutral to charged current events
in muon (anti)neutrino -- nucleon scattering and from these has
obtained values of effective coupling parameters $g_L^2=0.30005 \pm
0.00137$ and $g_R^2=0.03076 \pm 0.00110$ \cite{LlewellynSmith:ie}.
Standard Model (SM) predictions of these parameters based on a global
fit to non-NuTeV data, cited as $[g_L^2]_\mathrm{SM}=0.3042$ and
$[g_R^2]_\mathrm{SM}=0.0301$ in Ref.~\cite{Zeller:2001hh}, differ from
the NuTeV result by $3\sigma$ in $g_L^2$.  The significance of the
result remains controversial \cite{Davidson:2001ji} and a critical
examination of the initial analysis is ongoing, but it remains a
distinct possibility that the discrepancy with the SM prediction is
genuine and that its resolution lies in physics beyond the SM
\cite{Chanowitz:2002cd}.

Neglecting $g_R^2$, the ratio of neutral to charged current events is
simply $g_L^2$.  Since the NuTeV value for $g_L^2$ is \textit{smaller}
than its SM prediction, possible \textit{new} physics explanations of
the NuTeV anomaly would be those that suppress the neutral current
cross sections over the charged current cross sections, or enhance the
charged current cross sections over the neutral current cross
sections.  Two classes of models have been proposed which accomplish
this.

The first class comprises models which introduce new neutrino-quark
interactions, mediated by leptoquarks or extra $U(1)$ gauge bosons
($Z'$'s), which interfere either destructively with the $Z$-exchange
amplitude, or constructively with the $W$-exchange amplitude
\cite{Davidson:2001ji, Ma:2001md}.  Models in this class are
constrained strongly by lepton universality and predict gauge boson
masses in the several 100~GeV to TeV range, within reach of LHC.
Models of the second class suppress $Z\nu\nu$ and $W\mu\nu_\mu$
couplings by mixing the neutrino with heavy gauge singlet states
(neutrissimos) \cite{numix,Chang:1994hz,LOTW1,LORTW2}.  Suppressions
of the neutrino-gauge couplings also affect most other electroweak
observables and may violate lepton universality.  These models predict
new heavy particles which might be found at LHC and can be constrained
by tests of lepton universality, lepton flavor violation \cite{mega,
  taumugamma,meg,meco}, muon $g-2$
\cite{Sichtermann:2003cc,Ma:2002df}, and violations of CKM unitarity
\cite{Langacker:2003tv}.

\subsection{Lorentz noninvariance, CPT violation and decoherence}

CPT is a symmetry in any theory that satisfies the three assumptions
that are normally taken for granted: (1) locality, (2) Lorentz
invariance, and (3) hermiticity of the Hamiltonian.  In particular, it
predicts that the mass is common for a particle and its anti-particle.
Any violation of CPT would have profound consequences on
fundamental physics.

The best limit on CPT violation is in the neutral kaon system,
$|m(K^0) - m(\overline{K}^0)| < 10^{-18} m_K = 0.50 \times
10^{-18}$~GeV \cite{Hagiwara:fs}.  Such a stringent bound does
not seem to naively allow sizable CPT violation in the neutrino sector.
However, the kinematic parameter is mass-squared instead of mass, and
the constraint may naturally be considered on the CPT-violating
difference in mass-squared $|m^2(K^0) - m^2(\overline{K}^0)| <
0.25$~eV$^2$.  In comparison, the combination of SNO and KamLAND data
leads to the constraint $|\Delta m^2_\nu - \Delta m^2_{\bar{\nu}}| <
1.3 \times 10^{-3}$~eV$^2$ (90\% CL) and hence currently the best
limit on CPT violation \cite{Murayama:2003zw}.

New motivation for considering CPT violation among neutrinos arose
recently with the observation that the LSND, solar, and atmospheric
neutrino data can be accommodated simultaneously without invoking a
sterile neutrino provided the neutrino and anti-neutrino masses are
not the same violating CPT
\cite{Murayama:2000hm,Barenboim:2001ac,ales}.  

\begin{figure}[tbp]
  \centering
  \includegraphics[width=0.4\textwidth]{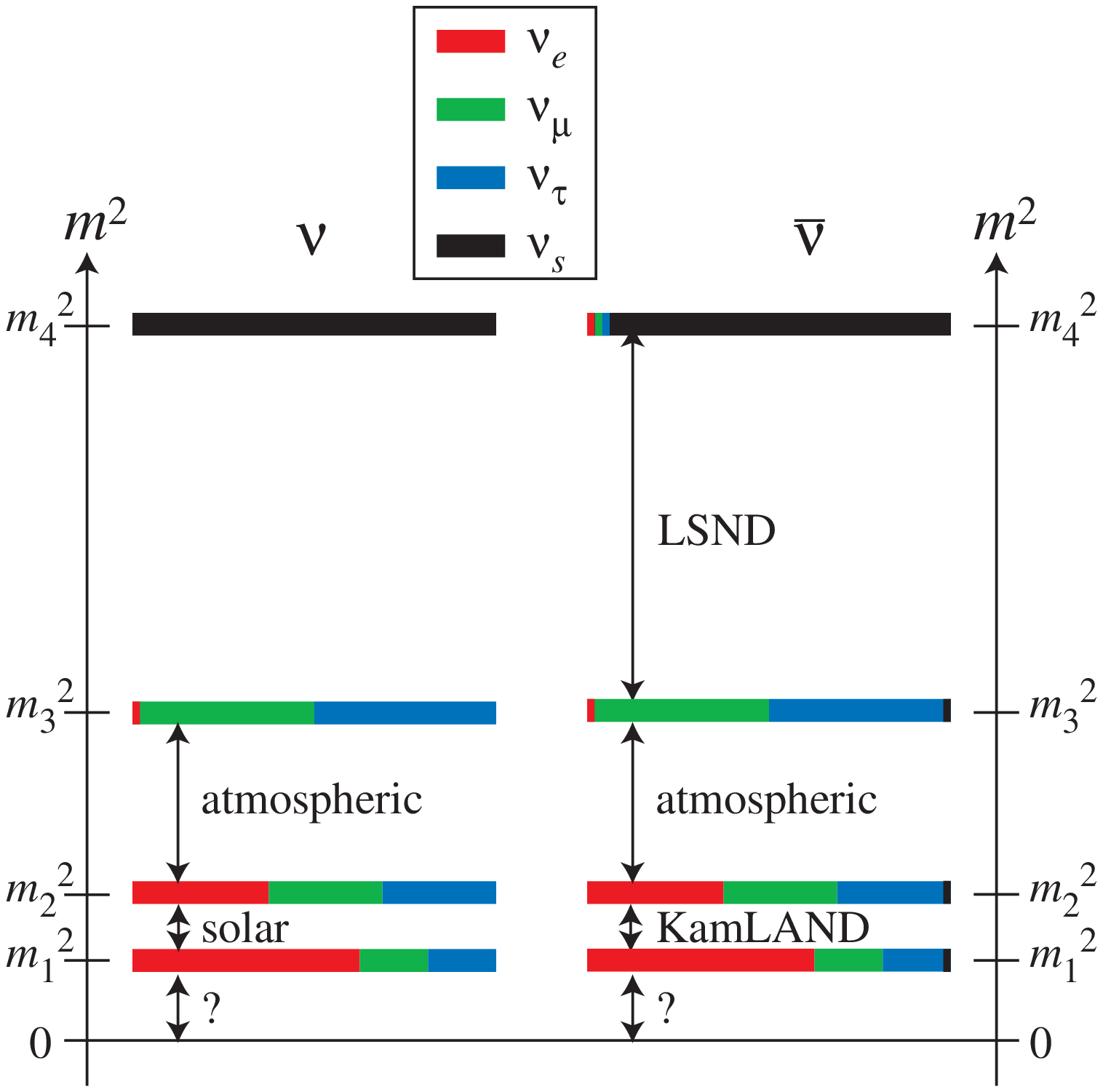}
  \includegraphics[width=0.4\textwidth]{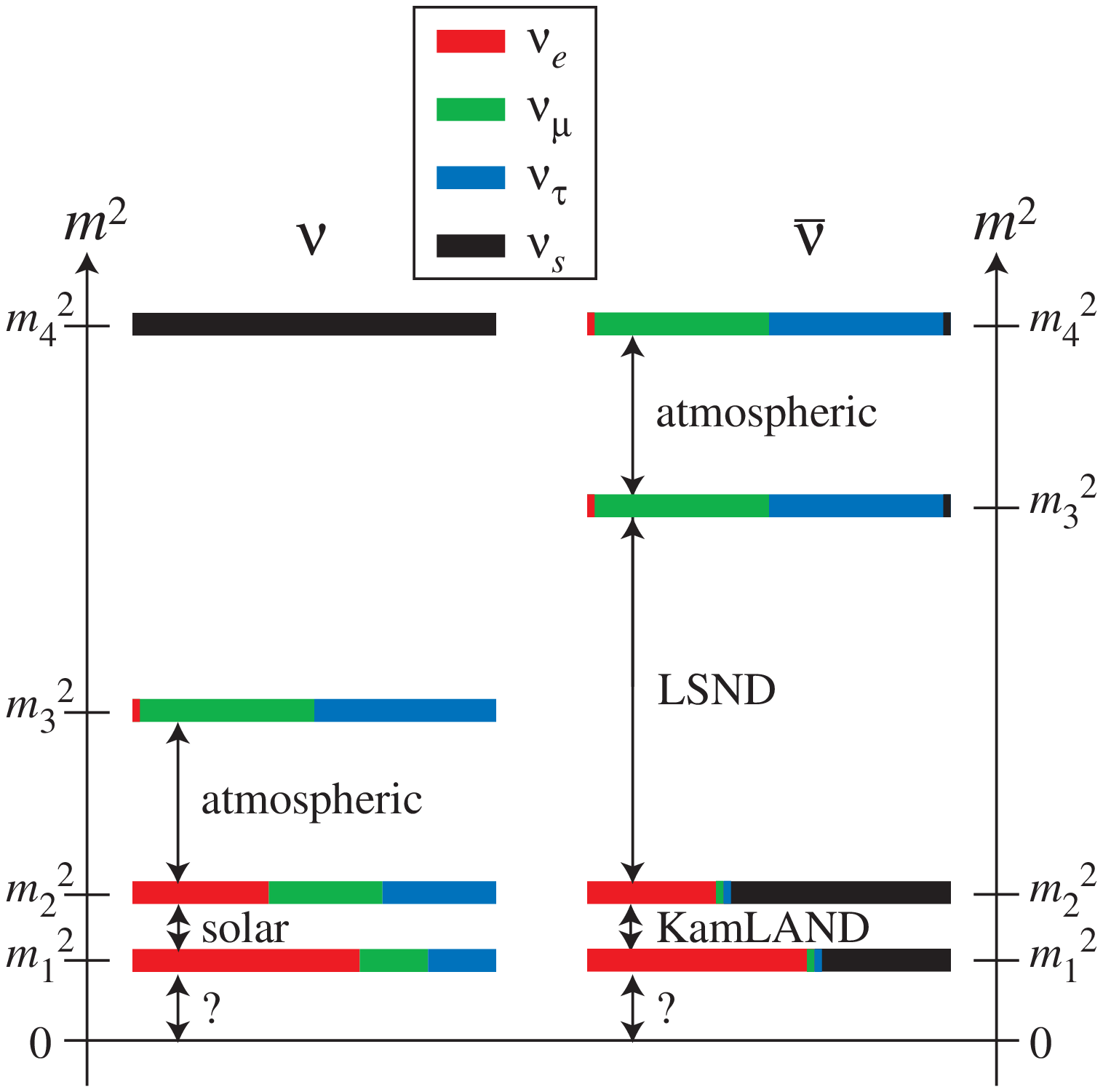}
  \caption{The revised proposal in \cite{Barger:2003xm} that combines
    CPT violation and a sterile neutrino.  The neutrinos always have
    $2+2$ spectrum, while the anti-neutrinos may have either $3+1$ or
    $2+2$ spectrum.}
  \label{fig:newCPTviolation}
\end{figure}

The KamLAND data, however, require $\bar{\nu}_e \rightarrow
\bar{\nu}_{\mu,\tau}$ oscillations with parameters consistent with the
solar neutrino oscillation, and CPT-violation alone cannot explain
LSND.  A different proposal to explain LSND and atmospheric
anti-neutrino oscillations with a single $\Delta m^2$
\cite{Barenboim:2002rv}, is excluded by the atmospheric neutrino data
\cite{Gonzalez-Garcia:2003jq}.  
The introduction of CPT violation improves significantly four neutrino
(three active and one sterile) fits to all neutrino data (including LSND)
\cite{Barger:2003xm}.  This is due to the fact that the short-baseline
experiments constraining the interpretation of the LSND data with
a sterile neutrino involve mostly neutrinos but not anti-neutrinos,
and the $3+1$ spectrum (Fig.~\ref{fig:newCPTviolation}) 
is allowed if there is little mixing of the sterile
state with the active ones.

Other possibilities that go beyond conventional quantum
field theory have been proposed as a way to understand the LSND
anomaly. Decoherence is one such possibility \cite{decoherence}, which
can be tested using neutrino oscillation studies.


\section{Conclusion}

In this report, we presented a brief review of the present
knowledge of neutrino physics and what we can learn from planned
experiments in the next decade. The discussion group feels that three 
three most important experiments (beyond the KATRIN experiment which
is already under way) that will
have a significant impact on clarifying the nature of neutrino mass
hierarchy, the nature of the neutrino (Dirac or Majorana) as well as the
search for physics beyond the Standard Model are: (i) search for
$\beta\beta_{0\nu}$ decay, (ii) determination of the sign of the $\Delta
m^2_{13}$ and (iii) measurement of the value of $\theta_{13}$. The
last one  will not only specify the neutrino mass matrix more precisely
than we know today but will considerably narrow the field of models.
Next in our top priority list is the measurement of the Dirac phase, which
will give a partial understanding of CP violation in the leptonic sector.

We believe that all support should be given to MiniBooNE experiment till
it provides a complete resolution of the LSND result. If the MiniBooNE
confirms the LSND result, we need to completely revise our current
understanding of neutrinos and, perhaps, particle physics.  Therefore
 search for properties of the sterile neutrinos
will become a very high priority item at the same level as those discussed
in the previous paragraph.

 Within the three neutrino picture, the precise measurement
of the solar and atmospheric mixing angles will significantly help
discriminate among various new physics possibilities. We consider it as
the next level of priority.

If MiniBooNE does not confirm LSND, the light sterile neutrinos could still
be playing a subdominant role in solar neutrino physics, as has been
suggested by several theoretical models.
At the next level of priority, we consider items such as: (i)search for
subdominant effects of light sterile neutrinos using precision
measurements of pp neutrinos from the Sun; (ii) search for neutrino
magnetic moment, whose values below the
current astrophysical limit of $10^{-11}\mu_B$ 
will be a sure indication of TeV scale new physics, such as a TeV scale
left-right model, horizontal models, or large extra dimensions; (iii)
searches for exotic physics involving neutrinos that can test the limits
of the assumptions on which the Standard Model is based 
e.g., the violation of Lorentz invariance, the existence of new long range
forces coupled to lepton number, CPT violation, etc.


\begin{thebibliography}{99}


\def\plb#1#2#3{Phys.\ Lett.\       {\bf B#1}, #2  (#3)}
\def\npb#1#2#3{Nucl.\ Phys.\       {\bf B#1}, #2  (#3)}
\def\prd#1#2#3{Phys.\ Rev.\        {\bf D#1}, #2  (#3)}
\def\prl#1#2#3{Phys.\ Rev.\ Lett.\ {\bf #1},  #2  (#3)}
\def\mpl#1#2#3{Mod.\ Phys.\ Lett.\ {\bf A#1}, #2  (#3)}
\def\rep#1#2#3{Phys.\ Rep.\        {\bf #1}, #2   (#3)}
\def\sci#1#2#3{Science             {\bf #1}, #2   (#3)}
\def\astro#1#2#3{Astrophys.\ J.\   {\bf #1}, #2   (#3)}
\def\epj#1#2#3{Eur.\ Phys.\ J.\   {\bf C#1}, #2   (#3)}
\def\jhep#1#2#3{JHEP              {\bf #1}, #2   (#3)}
\def\ptp#1#2#3{Prog.\ Theor.\ Phys.\ {\bf #1}, #2  (#3)}
 
\bibitem{BPont57} B.~Pontecorvo, 
                  Zh.\ Eksp.\ Teor.\ Fiz.\ {\bf 33} (1957) 549 
                and {\bf 34} (1958) 247;
Z.~Maki, M.~Nakagawa and S.~Sakata, 
Prog.\ Theor.\ Phys.\  {\bf 28} (1962) 870;
 B. Pontecorvo, 
Zh. Eksp. Teor. Fiz. {\bf 53} (1967) 1717.

\bibitem{barger} V. Barger, K. Whisnant and D. Marfatia,
Int.J.Mod.Phys. {\bf E12}, 569 (2003); C. Gonzales-Garcia and Y. Nir,
Rev.Mod.Phys. {\bf 75}, 345 (2003); A. Smirnov, hep-ph/0311259; S. Pakvasa
and J. W. F. Valle, hep-ph/0301061; S.~M.~Bilenky,
C.~Giunti and W.~Grimus,
Prog.\ Part.\ Nucl.\ Phys.\ {\bf 43} (1999) 1; S. F. King,
Rept.Prog.Phys. {\bf
67}, 107  (2004); G. Altarelli and F. Feruglio, hep-ph/0405048;
B.~Bajc, F.~Nesti, G.~Senjanovic and F.~Vissani,
{\it Proceedings of 17th Rencontres de Physique de la Vallee d'Aoste},
La Thuile, 9-15 Mar 2003,  M. Greco ed., page 103-143; R. N. Mohapatra,
hep-ph/0211252 (to appear in NJP,2004).


       

\bibitem{Valle} S.~M.~Bilenky {\it et al.},
              Phys.\ Lett.\  B{\bf 94} (1980) 495; J. Schechter and
J.~W.~F.~Valle, {Phys. Rev.} {\bf D22} (1980) 2227; M.~Doi {\it et al.},
Phys.\ Lett.\  \textbf{B102} (1981) 323.

\bibitem{moh1}  R. N. Mohapatra and J. Vergados, Phys. Rev. Lett. {\bf 47}, 1713 (1981);
R. N. Mohapatra, Phys. Rev. {\bf D 34} (1986) 3457;
B. Brahmachari and  E. Ma,  Phys.Lett. {\bf B536} (2002) 259.



\bibitem{klapdor} H.~V.~Klapdor-Kleingrothaus, A.~Dietz, H.~L.~Harney
and I.~V.~Krivosheina, Mod. Phys. Lett. {\bf A16} (2001) 2409, hep-ph/0201231;
H.~V.~Klapdor-Kleingrothaus, I.~V.~Krivosheina, A.~Dietz and O.~Chkvorets,
Phys.\ Lett.\ {\bf B586} (2004) 198.



\bibitem{hannestad} S. Hannestad, hep-ph/0310220.


\bibitem{boris} B. Kayser, in {\it CP violation}, ed. C. Jarlskog (World
Scientific, 1988);  S. Pascoli, S. T. Petcov, L. Wolfenstein, 
Phys.Lett. {\bf B524}, 319 (2002);  
Z-Z. Xing,
hep-ph/0307359; A. Broncano, M.B. Gavela, E. Jenkins, Nucl.Phys. {\bf
B672},  163 (2003).

\bibitem{boris_2} A. de Gouv\^ea, B. Kayser and R. N. Mohapatra,
Phys.Rev. {\bf D67},053004 (2003).


\bibitem{minakata}  H.~Minakata, H.~Nunokawa, S.~Parke
hep-ph/0208163; Phys.Rev. {\bf D66}, 093012 (2002);  J. Burguet-Castell {\it et al.}, Nucl.Phys. B646
(2002) 301 (2002); S. Pascoli, S.T. Petcov,
W. Rodejohann; Phys.Rev. {\bf D68}, 093007 (2003); H. Minakata,
hep-ph/0402197 and references therein.

\bibitem{caldwell} D. Caldwell and R. N. Mohapatra, Phys. Rev. {\bf D 46},
3259 (1993); J. Peltoniemi and J. W. F. Valle, Nucl. Phys. {\bf B 406},
409 (1993); J. Peltoniemi, D. Tommasini and J. W. F. Valle,
Phys. Lett. {\bf B 298}, 383 (1993).

\bibitem{other} S. Bilenky, W. Grimus, C. Giunti and
T. Schwetz, hep-ph/9904316;  V. Barger {\it et al.}, Phys. Lett. {\bf B 489}, 345 (2000); for a review, see
S. Bilenky, C. Giunti and W. Grimus,  Prog.Part.Nucl.Phys. {\bf 43}, 1
(1999). 
 

\bibitem{sorel} M. Sorel, J. Conrad and M. Shavitz, hep-ph/0305255.

\bibitem{raffelt} See G. Raffelt, {\it Stars as Laboratories for
Fundamental Physics}, Chicago University Press (1996).

\bibitem{cmp} Y. Chikashige, R. N. Mohapatra and R. D. Peccei,
Phys. Lett. {\bf B 98}, 265 (1981).

\bibitem{beacom} J.F.~Beacom {\it et al.,} hep-ph/0309267.

 
 

\bibitem{Yanagida:1980}
T.~Yanagida, in \emph{Proceedings of the Workshop on the Unified Theory
and the
  Baryon Number in the Universe} (O.~Sawada and A.~Sugamoto, eds.), KEK,
  Tsukuba, Japan, 1979, p.~95;
M.~Gell-Mann, P.~Ramond, and R.~Slansky, \emph{Complex spinors and unified
  theories}, in \emph{Supergravity} (P.~van Nieuwenhuizen and
D.~Z. Freedman,
  eds.), North Holland, Amsterdam, 1979, p.~315;
S.~L. Glashow, \emph{The future of elementary particle physics}, in
  \emph{Proceedings of the 1979 Carg{\`e}se Summer Institute on Quarks and
  Leptons} (M.~L{\'e}vy, J.-L. Basdevant, D.~Speiser, J.~Weyers,
R.~Gastmans,
  and M.~Jacob, eds.), Plenum Press, New York, 1980, pp.~687--713;
R.~N. Mohapatra and G.~Senjanovi{\'c}, Phys. Rev. Lett. \textbf{44}
(1980),
  912.

\bibitem{reconstruction} see, for example, M.~Frigerio and A.Yu.~Smirnov,
Nucl.\ Phys.\ B {\bf 640}, 233 (2002);
Phys.\ Rev.\ D {\bf 67}, 013007 (2003).

\bibitem{noMaj} V.~Barger, S.L.~Glashow, P.~Langacker and D.~Marfatia,
Phys.\ Lett.\ B {\bf 540}, 247 (2002);
 S.~Pascoli, S.~T.~Petcov and W.~Rodejohann,
Phys.\ Lett.\ B {\bf 549}, 177 (2002).

\bibitem{theta_23_andre} A.~de Gouv\^ea,
hep-ph/0401220.



\bibitem{PPaddendum}  F.~Vissani,
JHEP {\bf 9906} (1999) 022; F.~Feruglio, A.~Strumia and F.~Vissani,
Nucl.\ Phys.\ B {\bf 637} (2002) 345; S. Pascoli and S.T. Petcov,
{\em Phys. Lett.} {\bf B 580} (2004) 280.

\bibitem{shrock} R. N. Mohapatra and J. W. F. Valle, Phys. Rev. {\bf D
34} (1986) 1642; T. Appelquist, M. Piai, and R. Shrock,
Phys. Rev. D {\bf 69},
  015002 (2004).]


\bibitem{so10} K. S. Babu and R. N. Mohapatra, Phys. Rev. Lett. {\bf 70}
(1993) 2845;
K.~Matsuda, Y.~Koide, T.~Fukuyama and H.~Nishiura,
Phys. Rev. D {\bf 65} (2002) 033008;
B.~Bajc, G.~Senjanovic and F.~Vissani,
Phys.\ Rev.\ Lett.\  {\bf 90} (2003) 051802;
H.~S.~Goh, R.~N.~Mohapatra and S.~P.~Ng,
Phys.\ Lett.\ B {\bf 570} (2003) 215;
M. C. Chen and K. T. Mahanthappa, Phys. Rev. {\bf D 62} (2000)
113007.

\bibitem{so1016}  K. S. Babu, J. C. Pati and
F. Wilczek, Nucl. Phys. {\bf B566} (2000) 33;
 C. Albright and S. M. Barr,
Phys. Rev. Lett. {\bf 85} (2001) 244; 
T. Blazek, S. Raby and K. Tobe, Phys. Rev. {\bf D62} (2000) 055001;
 Z. Berezhiani and A. Rossi, Nucl. Phys. {\bf B594} (2001) 113; 
G.~G.~Ross and L.~Velasco-Sevilla, Nucl.\ Phys.\ B {\bf 653} (2003) 3.

\bibitem{Chankowski:1993tx}
P.~H. Chankowski and Z.~Pluciennik, Phys. Lett. \textbf{B316} (1993)
312;
K.~S. Babu, C.~N. Leung and J.~Pantaleone, Phys. Lett. \textbf{B319}
(1993) 191; 
S.~Antusch, M.~Drees, J.~Kersten, M.~Lindner and M.~Ratz, Phys. Lett.
\textbf{B519} (2001) 238; 
Phys.\ Lett.\ B {\bf 525} (2002) 130.

\bibitem{Casas:1999tg}
J.~A. Casas, J.~R. Espinosa, A.~Ibarra and I.~Navarro, Nucl. Phys.
\textbf{B573} (2000) 652;  
P.~H. Chankowski and S.~Pokorski, Int. J. Mod. Phys. \textbf{A17} (2002)
575;\\ 
S.~Antusch, J.~Kersten, M.~Lindner and M.~Ratz, Nucl. Phys. \textbf{B674}
(2003) 401.
 
\bibitem{balaji} K.~R.~S. Balaji, A.~S. Dighe, R.~N. Mohapatra, and
M.~K. Parida,  Phys. Rev. Lett. \textbf{84} (2000),
5034; R.~N. Mohapatra, M.~K. Parida, and G.~Rajasekaran, Phys.Rev. {\bf 
D69} (2004) 053007.

\bibitem{Antusch:2002fr}
S.~Antusch and M.~Ratz,
JHEP {\bf 0211}, 010 (2002).





\bibitem{FY}
M. Fukugita and T. Yanagida, \plb{174}{45}{1986}.

\bibitem{Fujii:2002jw}
M.~Fujii, K.~Hamaguchi and T.~Yanagida,
Phys.\ Rev.\ D {\bf 65}, 115012 (2002)

\bibitem{di2} W.~Buchm\"uller, P.~Di Bari and M.~Pl\"umacher,
Nucl.\ Phys.\ B {\bf 665}, 445 (2003).

\bibitem{strumia}
T.~Hambye {\it et al.},
hep-ph/0312203.


\bibitem{gian}
G.~F.~Giudice {\it et al.,}
hep-ph/0310123.

\bibitem{Buchmuller:2004nz}
W.~Buchm\"{u}ller, P.~Di Bari and M.~Pl\"{u}macher,
hep-ph/0401240.

\bibitem{antusch:2004xy}
S.~Antusch and S.~F.~King,
hep-ph/0405093.



\bibitem{dip}
S.~Davidson,
JHEP {\bf 0303} (2003) 037;
S.~Davidson and A.~Ibarra,
 Nucl.\ Phys.\ B {\bf 648} (2003) 345

\bibitem{afsmirnov}
E.~K.~Akhmedov, M.~Frigerio and A.~Y.~Smirnov,
JHEP {\bf 0309} (2003) 021.


\bibitem{apostolos}
A.~Pilaftsis,
Phys.\ Rev.\ D {\bf 56} (1997) 5431; Nucl.\ Phys.\ B {\bf 504} (1997) 61.

\bibitem{Branco:2001pq}
G.~C.~Branco, T.~Morozumi, B.~M.~Nobre and M.~N.~Rebelo,
Nucl.\ Phys.\ B {\bf 617} (2001) 475

\bibitem{BP}
W.~Buchm\"uller and M.~Pl\"umacher,
Int.\ J.\ Mod.\ Phys.\ A {\bf 15} (2000) 5047

\bibitem{Buchmuller:2002rq}
W.~Buchm\"uller, P.~Di Bari and M.~Pl\"umacher,
Nucl.\ Phys.\ B {\bf 643} (2002) 367.






\bibitem{O'Donnell:1994am}
P.~J.~O'Donnell and U.~Sarkar,
Phys.\ Rev.\ D {\bf 49}, 2118 (1994)
[arXiv:hep-ph/9307279].

\bibitem{Hambye:2003ka}
T.~Hambye and G.~Senjanovic,
Phys.\ Lett.\ B {\bf 582}, 73 (2004);
A.~S.~Joshipura and E.~A.~Paschos,
hep-ph/9906498;
A.~S.~Joshipura, E.~A.~Paschos and W.~Rodejohann,
Nucl.\ Phys.\ B {\bf 611}, 227 (2001);
A.~S.~Joshipura, E.~A.~Paschos and W.~Rodejohann,
JHEP {\bf 0108}, 029 (2001);
W.~Rodejohann,
Phys.\ Lett.\ B {\bf 542}, 100 (2002).

\bibitem{FPSCRV}M.~Flanz, E.~A.~Paschos and U.~Sarkar,
Phys.\ Lett.\ B {\bf 345} (1995) 248
[Erratum-ibid.\ B {\bf 382} (1996) 447];
L.~Covi, E.~Roulet and F.~Vissani,
Phys.\ Lett.\ B {\bf 384} (1996) 169.


\bibitem{APTU}A.~Pilaftsis and T.~E.~J.~Underwood, hep-ph/0309342.


\bibitem{mv} R.~N.~Mohapatra and J.~W.~F.~Valle,
Phys.\ Rev.\ D {\bf 34}, 1642 (1986).


\bibitem{AB}C.~H.~Albright and S.~M.~Barr,
hep-ph/0312224.

\bibitem{DGRGKNR}G.~D'Ambrosio, G.~F.~Giudice and M.~Raidal,
Phys.\ Lett.\ B {\bf 575}, 75 (2003);
Y.~Grossman, T.~Kashti, Y.~Nir and E.~Roulet,
Phys.\ Rev.\ Lett.\  {\bf 91}, 251801 (2003).

\bibitem{THJMRSW}T.~Hambye, J.~March-Russell and S.~M.~West,
arXiv:hep-ph/0403183.

\bibitem{Dick:1999je}
K.~Dick, M.~Lindner, M.~Ratz and D.~Wright,
Phys.\ Rev.\ Lett.\  {\bf 84}, 4039 (2000);\\
H.~Murayama and A.~Pierce,
Phys.\ Rev.\ Lett.\  {\bf 89}, 271601 (2002).






\bibitem{superK} A. Joshipura and S. Mohanty,  Phys. Rev.
{\bf D66}, 012003 (2003).

\bibitem{munu} Z. Paraktchieva et al [MUNU collaboration], Phys.
Lett. {\bf B564} 190 (2003).

\bibitem{texono} H.~B.~Li et al [Texono collaboration], Phys. Rev. Lett. 
{\bf 90} (2003).

\bibitem{sno} S.~N.~Ahmed et al [SNO collaboration], nucl-ex/0309004.


\bibitem{grifols} A.~J.~Grifols, E.~Masso and S.~Mohanty, hep-ph/0401144.


\bibitem{bahcall} J.~N.~Bahcall and M.~H.~Pinsonneault, astro-ph/0402114.





\bibitem{bm}  F. Borzumati and A. Masiero, Phys. Rev. Lett. {\bf 57} 961
(1986); S. Lavignac, I. Masina and C.A. Savoy, Phys. Lett. {\bf B 520} 269
(2001); for recent review and references, see A. Masiero,
S. K. Vempati and O. Vives, hep-ph/0405017.


\bibitem{apo} A. Ilakovac and A. Pilaftsis, Nucl.\ Phys.\ B {\bf 437}
(1995) 491.



\bibitem{Grossman:2003gq}
Y. Grossman and S. Rakshit, hep-ph/0311310.

\bibitem{Banks:1995by}
T.~Banks, Y.~Grossman, E.~Nardi and Y.~Nir,
Phys.\ Rev.\ D {\bf 52}, 5319 (1995).

\bibitem{kuchi} R.~Kuchimanchi and R.~N.~Mohapatra,
Phys.\ Rev.\ D {\bf 48}, 4352 (1993).

\bibitem{Arkani-Hamed:2000bq}
N.~Arkani-Hamed {\it et al.,}
Phys.\ Rev.\ D {\bf 64}, 115011 (2001).


\bibitem{Arkani-Hamed:2000kj}
N.~Arkani-Hamed {\it et al.,}
hep-ph/0007001.


\bibitem{Borzumati:2000ya}
F.~Borzumati, K.~Hamaguchi, Y.~Nomura and T.~Yanagida,
hep-ph/0012118.

\bibitem{Grossman:1997is}
Y.~Grossman and H.~E.~Haber,
Phys.\ Rev.\ Lett.\  {\bf 78}, 3438 (1997).



\bibitem{KK} T. Kaluza, Sitzungsber.\ d.\ Preuss.\ Akad.\ d.\ Wiss.\
  Berlin, (1921) 966; O. Klein, Z.\ Phys.\ {\bf 37} (1926) 895.


\bibitem{ADD} N. Arkani-Hamed, S. Dimopoulos and G. Dvali, Phys.\ 
  Lett.\ {\bf B429} (1998) 263; I.~Antoniadis, N. Arkani-Hamed, S.
  Dimopoulos and G. Dvali, Phys.\ Lett.\ {\bf B436} (1998) 257.

\bibitem{DDG1} K.R. Dienes, E.  Dudas and T. Gherghetta, Phys.\ Lett.\
  {\bf B436} (1998) 55; Nucl.\ Phys.\ {\bf B537} (1999) 47.


\bibitem{DDG2} K.R. Dienes, E. Dudas and T. Gherghetta, Nucl.\ Phys.\ 
{\bf B557} (1999) 25.


\bibitem{DS} G. Dvali and A. Yu.\ Smirnov, Nucl.\ Phys.\ {\bf B563}
  (1999) 63.


\bibitem{ng} G. Mclaughlin and J. N. Ng, Phys. Rev. {\bf D 63}, 053002
(2001).

\bibitem{ynm} H. Yu, S.-P. Ng and R. N. Mohapatra, hep-ph/0404274.

\bibitem{cmy} D.O. Caldwell, R.N. Mohapatra and
 S.J. Yellin, Phys.\ Rev.\ {\bf D64} (2001) 073001.

\bibitem{apo1} A. Ioannisian and A. Pilaftsis, Phys.\ Rev.\ {\bf D62}
 (2000) 066001.


\bibitem{Zeller:2001hh}
[NuTeV Collaboration]
G.~P.~Zeller {\it et al.},
Phys.\ Rev.\ Lett.\  {\bf 88}, 091802 (2002);
Phys.\ Rev.\ D {\bf 65}, 111103 (2002);
K.~S.~McFarland {\it et al.}, hep-ex/0205080;
G.~P.~Zeller {\it et al.}, hep-ex/0207052.

\bibitem{LlewellynSmith:ie}
C.~H.~Llewellyn Smith, 
Nucl.\ Phys.\ B {\bf 228}, 205 (1983).


\bibitem{Davidson:2001ji}
S.~Davidson {\it et al.,}
JHEP {\bf 0202}, 037 (2002);
S.~Davidson,
hep-ph/0209316;
P.~Gambino,
hep-ph/0211009.

\bibitem{Chanowitz:2002cd}
M.~S.~Chanowitz,
Phys.\ Rev.\ D {\bf 66}, 073002 (2002).

\bibitem{Ma:2001md}
E.~Ma, D.~P.~Roy and S.~Roy,
Phys.\ Lett.\ B {\bf 525}, 101
(2002); 
E.~Ma and D.~P.~Roy, 
Phys.\ Rev.\ D {\bf 65}, 075021 (2002); 
Nucl.\ Phys.\ B {\bf 644}, 290 (2002). 


\bibitem{numix}
M.~Gronau, C.~N.~Leung and J.~L.~Rosner,
Phys.\ Rev.\ D {\bf 29}, 2539 (1984);
J.~Bernabeu {\it et al.,}
Phys.\ Lett.\ B {\bf 187}, 303 (1987);
K.~S.~Babu, J.~C.~Pati and X.~Zhang,
Phys.\ Rev.\ D {\bf 46}, 2190 (1992);
W.~J.~Marciano,
Phys.\ Rev.\ D {\bf 60}, 093006 (1999);
K.~S.~Babu and J.~C.~Pati,
hep-ph/0203029.

\bibitem{Chang:1994hz}
L.~N.~Chang, D.~Ng and J.~N.~Ng,
Phys.\ Rev.\ D {\bf 50}, 4589 (1994);

\bibitem{LOTW1}
W.~Loinaz, N.~Okamura, T.~Takeuchi and L.~C.~R.~Wijewardhana,
Phys.\ Rev.\ D {\bf 67}, 073012 (2003);
T.~Takeuchi,
hep-ph/0209109;
T.~Takeuchi, W.~Loinaz, N.~Okamura and L.~C.~R.~Wijewardhana,
hep-ph/0304203.

\bibitem{LORTW2}
W.~Loinaz {\it et al.,}
Phys.\ Rev.\ D {\bf 68}, 073001 (2003).



\bibitem{mega}
M.~L.~Brooks {\it et al.}  [MEGA Collaboration],
Phys.\ Rev.\ Lett.\  {\bf 83}, 1521 (1999). 

\bibitem{taumugamma}
K.~Abe {\it et al.}  [Belle Collaboration],
arXiv:hep-ex/0310029;
C.~Brown  [BABAR Collaboration],
eConf {\bf C0209101}, TU12 (2002)
[Nucl.\ Phys.\ Proc.\ Suppl.\  {\bf 123}, 88 (2003)].


\bibitem{meg}
S.~Ritt  [MUEGAMMA Collaboration],
Nucl.\ Instrum.\ Meth.\ A {\bf 494} (2002) 520.
See also the MEG Collaboration website at \texttt{http://meg.web.psi.ch/}.

\bibitem{meco}
J.~L.~Popp  [MECO Collaboration],
Nucl.\ Instrum.\ Meth.\ A {\bf 472}, 354 (2000);
M.~Hebert  [MECO Collaboration],
Nucl.\ Phys.\ A {\bf 721}, 461 (2003).

\bibitem{Sichtermann:2003cc}
E.~Sichtermann  [g-2 Collaboration],
eConf {\bf C030626}, SABT03 (2003)
[hep-ex/0309008].

\bibitem{Ma:2002df}
E.~Ma and D.~P.~Roy,
Phys.\ Rev.\ D {\bf 65}, 075021 (2002);
K.~S.~Babu and J.~C.~Pati,
Phys.\ Rev.\ D {\bf 68}, 035004 (2003);
T.~Fukuyama, T.~Kikuchi and N.~Okada,
Phys.\ Rev.\ D {\bf 68}, 033012 (2003).

\bibitem{Langacker:2003tv}
P.~Langacker,
AIP Conf.\ Proc.\  {\bf 698}, 1 (2004)
[hep-ph/0308145].



\bibitem{am} A.S.Joshipura and S. Mohanty , Phys. Lett. {\bf B584}
103 (2004)(arXiv:hep-ph/0310210).

\bibitem{grifols-masso} J.A.Grifols and E. Masso , Phys Lett {\bf
B 579}, 123(2004).



\bibitem{fornengo} 
N. Fornengo, M. Maltoni, R. Tomas Bayo, J. W. F. Valle, 
Phys. Rev. {\bf D 65} 013010 (2002).

\bibitem{atm3fam}
A.~Friedland, C.~Lunardini and M.~Maltoni,
hep-ph/0408264.

\bibitem{holanda2}
M. M. Guzzo, P. C. de Holanda, O. L. G. Peres, 
hep-ph/0403134.

\bibitem{pena1}
A. Friedland, C. Lunardini, C. Pe\~na-Garay, 
hep-ph/0402266.


\bibitem{bergmann1}
S. Bergmann, Y. Grossman, D. M. Pierce,
Phys. Rev. {\bf D61} 053005 (2000).

\bibitem{holanda3}
S. Bergmann, {\it et al.,}\/ Phys. Rev. {\bf D 62}, 073001 (2000).

\bibitem{rossi} 
Zurab Berezhiani, Anna Rossi, 
Phys. Lett. {\bf B 535},  207 (2002).

\bibitem{pena2}
S. Davidson, C. Pe\~na-Garay, N. Rius, A. Santamaria
JHEP 0303 (2003) 011

\bibitem{shrock1} K.~Fujikawa and R.~Shrock,
Phys.\ Rev.\ Lett.\  {\bf 45}, 963 (1980).

\bibitem{cmsv} M.~Cirelli, G.~Marandella, A.~Strumia and F.~Vissani,
hep-ph/0403158.

\bibitem{pl}
P.~Langacker,
Phys.\ Rev.\ D {\bf 58}, 093017 (1998)


\bibitem{bnv} V.~Berezinsky, M.~Narayan and F.~Vissani,
Nucl.\ Phys.\ B {\bf 658}, 254 (2003)

\bibitem{zr}
Z.~G.~Berezhiani and R.~N.~Mohapatra,
Phys.\ Rev.\ D {\bf 52}, 6607 (1995);
R.~Foot and R.~R.~Volkas, Phys.\ Rev.\ D {\bf 52}, 6595 (1995)
[arXiv:hep-ph/9505359].



\bibitem{alyos} K.~Benakli and A.~Y.~Smirnov,
Phys.\ Rev.\ Lett.\  {\bf 79}, 4314 (1997)

\bibitem{bgg} S.~M.~Bilenky, C.~Giunti and W.~Grimus,
Eur.\ Phys.\ J.\ C {\bf 1}, 247 (1998).

\bibitem{ales} A.~Strumia,
Phys.\ Lett.\ B {\bf 539}, 91 (2002).

\bibitem{bv} V.~S.~Berezinsky and A.~Vilenkin,
Phys.\ Rev.\ D {\bf 62}, 083512 (2000).



\bibitem{Hagiwara:fs} K.~Hagiwara {\it et al.}  [Particle Data Group
  Collaboration],
Phys.\ Rev.\ D {\bf 66}, 010001 (2002).

\bibitem{Murayama:2003zw}
H.~Murayama,
hep-ph/0307127.

\bibitem{Murayama:2000hm}
H.~Murayama and T.~Yanagida,
Phys.\ Lett.\ B {\bf 520}, 263 (2001).

\bibitem{Barenboim:2001ac}
G.~Barenboim, L.~Borissov, J.~Lykken and A.~Y.~Smirnov,
JHEP {\bf 0210}, 001 (2002).

\bibitem{Barenboim:2002rv}
G.~Barenboim, L.~Borissov and J.~Lykken,
Phys.\ Lett.\ B {\bf 534}, 106 (2002).

\bibitem{Gonzalez-Garcia:2003jq}
M.~C.~Gonzalez-Garcia, M.~Maltoni and T.~Schwetz,
Phys.\ Rev.\ D {\bf 68}, 053007 (2003).

\bibitem{Barger:2003xm}
V.~Barger, D.~Marfatia and K.~Whisnant,
Phys.\ Lett.\ B {\bf 576}, 303 (2003).

\bibitem{decoherence} J. Ellis, J. Hagelin, D.V. Nanopoulos and M. Srednicki,
Nucl. Phys. {\bf B241}, 381 (1984); E.~Lisi, A.~Marrone and D.~Montanino,
Phys.\ Rev.\ Lett.\  {\bf 85}, 1166 (2000)
[arXiv:hep-ph/0002053];
 G.~Barenboim and N.~Mavromatos, hep-ph/0404014.






\end{thebibliography}
\end{document}